%% file: IndexAndAnomalies.tex
\documentclass[12pt]{article}
\pdfoutput=1
\usepackage{jheppub}
\definecolor{cardinal}{rgb}{0.6,0,0}
\definecolor{darkgreen}{rgb}{0,0.5,0}
\definecolor{golden}{rgb}{0.92, 0.7, 0}
\definecolor{midnight}{rgb}{0, 0, 0.5}
\definecolor{darkblue}{rgb}{0.2, 0, 0.8}

\newcommand{\nn}{\nonumber}

\input{preamble}

\title{Supersymmetric Casimir Energy and the Anomaly Polynomial}
\author[1]{Nikolay Bobev}
\author[2]{Mathew Bullimore}
\author[3]{Hee-Cheol Kim}
\affiliation[1]{Instituut voor Theoretische Fysica, KU Leuven, Celestijnenlaan 200D, B-3001 Leuven, Belgium}
\affiliation[2]{Institute for Advanced Study, Einstein Dr., Princeton NJ 08450, USA}
\affiliation[3]{Perimeter Institute for Theoretical Physics, 31 Caroline St.\,N., Waterloo, ON N2L 6B9, Canada}

\emailAdd{nikolay@itf.fys.kuleuven.be}
\emailAdd{mathew.bullimore@gmail.com}
\emailAdd{heecheol1@gmail.com}

\abstract{We conjecture that for superconformal field theories in even dimensions, the supersymmetric Casimir energy on a space with topology $S^1\times S^{D-1}$ is equal to an equivariant integral of the anomaly polynomial. The equivariant integration is defined with respect to the Cartan subalgebra of the global symmetry algebra that commutes with a given supercharge. We test our proposal extensively by computing the supersymmetric Casimir energy for large classes of superconformal field theories, with and without known Lagrangian descriptions, in two, four and six dimensions.}

\begin{document}
\maketitle

\section{Introduction}
\label{sec:intro}

It is a well-known fact that in a two-dimensional CFT the Casimir energy on the cylinder is related to the conformal anomaly coefficient $c$. This is proven by performing a conformal transformation to flat space accompanied by the transformation law of the energy-momentum tensor. The Casimir energy can also be extracted from the partition function $Z$ on $S_\beta^1 \times S^1$ in the limit of infinite radius of the circle, $\beta \to \infty$, 
\be
Z \to e^{-\beta E} + \cdots\;, \qquad\qquad E = - \frac{c}{12}\; .
\label{eq:Glimit}
\ee
There have been attempts to generalize these results to CFTs in higher dimensions, see for example \cite{Herzog:2013ed}. However, there may be no general universal relation between Casimir energies and conformal anomalies in higher dimensions due to the existence of finite counterterms that render the result scheme dependent \cite{Assel:2015nca}.

The situation is more promising for SCFTs. For 4d $\cN=1$ SCFTs with a Lagrangian description, it was observed in \cite{Kim:2012ava,Assel:2014paa} that one can extract the conformal anomalies $a$ and $c$ from the partition function on $S_\beta^1 \times S^3$. The latter may be computed by supersymmetric localization. The result is
\be
Z = e^{-\beta E} I\;,
\label{eq:basic}
\ee 
where 
\be
\label{eq:Ebasic}
E =  \frac{2}{3}(a-c)(\omega_1+\omega_2)  + \frac{2}{27}(3c-2a) \frac{(\omega_1+\omega_2)^3}{\omega_1\omega_2} \; , \\[2mm]
\ee
and $I$ is the superconformal index~\cite{Romelsberger:2005eg,Kinney:2005ej}. The parameters $\omega_1$ and $\omega_2$ determine the geometry of $S^3$ and the background R-symmetry fields that must be turned on to preserve supersymmetry. 

The function $E$ gives the leading behavior of the partition function in the $\beta\to\infty$ limit, as in two dimensions. This result was further clarified in~\cite{Assel:2015nca,Assel:2014tba} where it was shown that there are no finite counterterms and $E$ is scheme-independent. The relation \eqref{eq:Ebasic} was further studied in \cite{Cassani:2014zwa} where the authors discussed a holographic interpretation of this result.\footnote{See also \cite{Spiridonov:2012ww,Ardehali:2013xya,Ardehali:2014zba,Ardehali:2014esa,Ardehali:2015hya,Lorenzen:2014pna,Nieri:2015yia} for related work on how the superconformal index or partition function of 4d $\mathcal{N} =1$ theories encodes various anomalies.} We refer to the quantity $E$ and its cousins for SCFTs in other even dimensions as the supersymmetric Casimir energy.

The purpose of this paper is to propose a simple universal formula for the supersymmetric Casimir energy $E$ in terms of the 't Hooft anomalies for continuous R-symmetry and flavor symmetries. Since conformal anomalies are related to R-symmetry anomalies by supersymmetry we will recover the result in \eqref{eq:Ebasic} in a limit. Specifically, we propose that the supersymmetric Casimir energy in $D$ (even) dimensions is an equivariant integral of the anomaly polynomial $A_{D+2}$\footnote{See \cite{Harvey:2005it} for a pedagogical exposition on anomalies and the anomaly polynomial.}, which we write schematically
\be
E_D = \int A_{D+2} \, .
\label{eq:conj}
\ee 
Here we work equivariantly with respect to a maximal torus of the global symmetry algebra commuting with the supercharges preserved by the partition function $Z$. The equivariant parameters are related to the expectation values of background vector multiplets for these symmetries, or equivalently the chemical potentials of the superconformal index $I$. This proposal is explained in  more detail in Section~\ref{sec:generalities}.

In Sections~\ref{sec:6d},~\ref{sec:4d} and \ref{sec:2d}, we perform numerous checks of our proposal for SCFTs with varying amounts of supersymmetry in two, four and six dimensions, with and without Lagrangian descriptions. We believe this provides ample evidence for our conjecture. We hope to return to proving the conjecture in future work.

We conclude in Section \ref{sec:discussion} with a summary and a collection of open questions. In Appendix \ref{app:1} we summarize some basic facts about equivariant differential forms and equivariant integration. Appendix \ref{app:2} is devoted to a discussion of the properties of various special functions that appear in our calculations.

\section{Generalities}
\label{sec:generalities}

The superconformal index of an SCFT in $D$ dimensions is defined as a trace over the Hilbert space $\cH$ in radial quantization,
\be
I(\beta_j)= \tr_{\cH} (-1)^F e^{-\gamma \{ Q, Q^\dagger \} } e^{-\sum_j \beta_j t_j} \;,
\ee
where $F$ is the fermion number, $Q$ is a supercharge, and $t_j$ are the generators of the Cartan subalgebra of the superconformal and flavor symmetry algebra commuting with $Q$, see~\cite{Romelsberger:2005eg,Kinney:2005ej}. The real parameters $\gamma$ and $\beta_j$ are called chemical potentials. By a standard argument, the superconformal index is independent of the parameter $\gamma$ and can be expressed as a trace over the subspace $\cH_Q \subset \cH$ of states saturating the unitarity bound $\{ Q , Q^\dagger \} \geq 0$, that is
\be
I(\beta_j) = \tr_{\cH_Q} (-1)^Fe^{-\sum_j \beta_j t_j} \, .
\ee
The superconformal index therefore receives contributions from short representations of the superconformal algebra that cannot combine into long representations. As a consequence, it is invariant under all deformations of the theory that preserve the supercharge $Q$, and in particular under marginal deformations of the fixed point. 

If the superconformal fixed point appears at the endpoint of a renormalization group flow triggered by a deformation of a free theory, the superconformal index can be evaluated in the free theory as a Plethystic exponential of the single-letter index. The Plethystic exponential of a function $f(x)$ with a Taylor series expansion around $x=0$, $f(x) =\sum_{n=0}^{\infty}a_nx^n$, is defined as
\begin{equation}
\text{PE}[ \, f(x) \, ] \equiv \text{exp}\left( \, \sum_{n=1}^{\infty}\frac{f(x^n)-f(0)}{n} \, \right) = \frac{1}{\prod_{n=1}^{\infty}(1-x^n)^{a_n}}\;,
\end{equation}
with an obvious generalization to functions of many variables. The single-letter index only receives contributions  from the elementary fields of the free theory and their derivatives. The Plethystic exponential then sums the contributions from all ``words" built out of the elementary fields. In a gauge theory, one should include only the contributions from gauge-invariant states. This can be accomplished by introducing additional chemical potentials for the gauge symmetry, which are then integrated over. The superconformal index can thus be viewed as a series expansion in $e^{-\beta_i}$.

A closely related object is the partition function of the theory on $S^1 \times S^{D-1}$ preserving the same supercharge $Q$. The details of the supersymmetric background will of course depend on the dimension $D$ and the amount of supersymmetry involved. Typically, the partition function depends on the radius $\beta$ of $S^1$ and a number of parameters $\mu_j$ describing the metric on $S^{D-1}$ and expectation values of background $R$-symmetry and flavor vector multiplets. The partition function $Z(\beta,\mu_j)$ can often be computed exactly by supersymmetric localization using the supercharge $Q$ and typically takes the form of a matrix integral of 1-loop determinants and in some cases non-perturbative contributions.

It is intuitively clear by cutting the path integral on $S^1$ that the supersymmetric partition function $Z(\beta,\mu_j)$ should be closely related to the superconformal index $I(\beta_j)$. Indeed, it has been demonstrated in a number of examples, that\footnote{With 4d $\cN=1$ supersymmetry, it was reported in~\cite{Assel:2014paa} that there could be a physically meaningful contribution to the exponential at order $\cO(\beta^{-1})$. However, it was subsequently explained \cite{Assel:2015nca} that this is absent when regularizing in a way that is compatible with the relevant supercharge $Q$. We expect similar statements in two and six dimensions. In any case, the presence of such terms would not affect our conjecture regarding the supersymmetric Casimir energy $E$, which is the coefficient of the $\cO(\beta)$ term.}
\be
Z (\beta, \mu_j) = e^{-\beta E(\mu_j)} I(\beta_j)\;,
\ee
where $\beta_j = \beta \mu_j$ and $E(\mu_j)$ is a finite Laurent polynomial in the rescaled chemical potentials $\mu_j$. The extraction of this result often requires careful regularization of 1-loop determinants and/or re-summation of infinite number of non-perturbative contributions to the localized path integral $Z(\beta,\mu_j)$.

The function $E(\mu_j)$ can be interpreted as a supersymmetric Casimir energy and should be physically meaningful. Indeed, given that the superconformal index $I(\beta_j)$ is a series expansion in $e^{-\beta \mu_j}$, it can be extracted from the supersymmetric partition function in the limit of infinite radius of $S^1$,
\be
E(\mu_j) = - \lim_{\beta \to \infty} \frac{\del}{\del\beta} \log Z(\beta,\mu_j) \, .
\ee
The supersymmetric Casimir energy $E(\mu_j)$ is a finite Laurent polynomial in the $\mu_j$, whose coefficients are particular linear combinations of the anomaly coefficients for conformal, $R$-symmetry, and flavor symmetries used in the construction of the partition function.\footnote{In supersymmetric theories the conformal anomalies are related by supersymmetry to $R$-symmetry anomalies.}

The purpose of this paper is to propose that the supersymmetric Casimir energy in even dimensions can be extracted directly from the anomaly polynomial of the theory. We conjecture that $E(\mu_j)$ is an equivariant integral of the anomaly polynomial $A_{D+2}$ over $\mathbb{R}^D$. We work equivariantly with respect to the Abelian symmetry group generated by the charges $t_j$ commuting with $Q$. The equivariant parameters are the corresponding chemical potentials $\mu_j$. We can write this as
\be
E(\mu_j) = \int_{\mu_j} A_{D+2} \, .
\ee
Note that for this conjecture to make sense we must view the anomaly polynomial $A_{D+2}$ as an equivariant characteristic class on $\mathbb{R}^D$. In equivariant cohomology, it is quite natural to have equivariant forms whose degrees are greater than the dimension of the manifold and whose equivariant integrals are non-zero. We refer the reader to Appendix~\ref{app:1} for a summary of equivariant characteristic classes and equivariant integration. Numerous examples will be considered below.

In the remaining sections, we will test this conjecture extensively for a number of SCFTs with and without Lagrangian descriptions in two, four and six dimensions.

\section{Six dimensions}
\label{sec:6d}

\subsection{$\cN = (2,0)$ supersymmetry}
\label{ssec:6d(2,0)}

The six-dimensional $(2,0)$ superconformal algebra is $\mathfrak{osp}(8^*|4)$. This superconformal algebra has the maximal bosonic subalgebra $\mathfrak{so}(2,6) \oplus \mathfrak{usp}(4)$. We denote the Cartan generators of the six-dimensional conformal algebra $\mathfrak{so}(2,6)$ by $(\Delta,h_1,h_2,h_3)$. The generator $\Delta$ corresponds to dilatations and $(h_1,h_2,h_3)$ to rotations in three orthogonal planes in $\mathbb{R}^6$. We denote the Cartan generators of the R-symmetry algebra $\mathfrak{usp}(4) = \mathfrak{so}(5)$ by $(r_1,r_2)$.

The supersymmetry generators can be labelled $Q_{h_1,h_2,h_3}^{r_1,r_2}$ with the indices taking the values $\pm \frac{1}{2}$. To simplify notation we will write $\pm$ instead. There are sixteen Poincar\'e supercharges consisting of the supercharges with $h_1h_2h_3<0$. The remaining sixteen supercharges with $h_1h_2h_3>0$ are the conformal supercharges. In radial quantization, conjugation reverses the sign of $h_1,h_2,h_3,r_1,r_2$ and so interchanges Poincar\'e and conformal supercharges.\footnote{We work in Euclidean signature and thus the conformal algebra is $\mathfrak{so}(1,7)$.}

The superconformal index in six dimensions was introduced in~\cite{Bhattacharya:2008zy}. Here, we will define the superconformal index using the supercharge $Q \equiv Q_{---}^{++}$. A different choice of supercharge will lead to an equivalent superconformal index. 
This supercharge generates an $\mathfrak{su}(1|1)$ subalgebra with
\be
\{ Q,Q^\dagger \} = \Delta - 2(r_1+r_2) - (h_1+h_2+h_3) \, .
\ee
The superconformal index counts states in short representations of the superconformal algebra annihilated by $Q$ and $Q^\dagger$, which therefore saturate the unitarity bound
\be
\Delta \ge 2(r_1+r_2) + h_1+h_2+h_3 \ .
\label{6dbound}
\ee
The superconformal index is defined by
\be
I = \tr_{\cH_Q} (-1)^F \prod_{j=1}^3 q_j^{h_j+\frac{r_1+r_2}{2}} p^{r_2-r_1} \ ,
\label{6dindexdef}
\ee
where $\cH_Q$ is the subspace of the Hilbert space in radial quantization that is annihilated by $Q$ and $Q^\dagger$. The four combinations $h_j + \frac{1}{2}(r_1+r_2)$ (with $j=1,2,3$) and $r_2-r_1$ form a basis for the space of linear combinations of Cartan generators commuting with $Q$. The corresponding fugacities are denoted $q_1$, $q_2$, $q_3$ and $p$. For convergence we assume that $|q_1|$, $|q_2|$, and $|q_3|<1$. $F$ is the fermion number, which we can define by $F=2h_1$.

\subsubsection{Tensor multiplet}

The tensor multiplet is a free theory consisting of a 2-form gauge field $B$ with self-dual curvature $H=dB=\star H$, fermions $\psi^{r_1r_2}_{h_1h_2h_3}$ with the same quantum numbers as the Poincar\'e supersymmetry generators with $h_1h_2h_3<0$ (and unrestricted values of $r_{1,2}$), and five real scalars $\phi$ in the fundamental representation of $\mathfrak{so}(5)$.

\begin{table}[h]
\begin{center}
  \begin{tabular}{ | c | c | c | c | c | c | c |}
    \hline
    $X$ & $h_1$ & $h_2$ & $h_3$ & $r_1$ & $r_2$ & \\ \hline
    $\phi$ & 0 & 0 & 0 & 1 & 0 & $p^{-1} \sqrt{q_1 q_2 q_3}$ \\ \hline
    $\phi$ & 0 & 0 & 0 & 0 & 1 & $p \sqrt{q_1 q_2 q_3}$ \\ \hline
    $\psi_{++-}^{++}$ & $\frac{1}{2}$ & $\frac{1}{2}$ & -$\frac{1}{2}$ & $\frac{1}{2}$ & $\frac{1}{2}$ & $-q_1q_2$ \\ \hline
    $\psi_{+-+}^{++}$ & $\frac{1}{2}$ & -$\frac{1}{2}$ & $\frac{1}{2}$ & $\frac{1}{2}$ & $\frac{1}{2}$ & $-q_1 q_3$ \\ \hline
    $\psi_{-++}^{++}$ & -$\frac{1}{2}$ & $\frac{1}{2}$ & $\frac{1}{2}$ & $\frac{1}{2}$ & $\frac{1}{2}$ & $-q_2 q_3$ \\ \hline
    $\del \psi $ & $\frac{1}{2}$ & $\frac{1}{2}$ & $\frac{1}{2}$ & $\frac{1}{2}$ & $\frac{1}{2}$ & $q_1q_2q_3$ \\ \hline
 \end{tabular}
\end{center}
\caption{The fields of the $(2,0)$ tensor multiplet saturating the bound~\eqref{6dbound} and their contributions to the superconformal index. Note that there is a contribution from a fermionic equation of motion, denoted schematically by $\partial \psi$. Recall also that $\Delta(\phi) = 2$, $\Delta(\psi) = 5 / 2$ and $\Delta(H)=3$.}
\label{free}
\end{table}

Since the tensor multiplet is a free theory, the superconformal index can be evaluated by enumerating contributions to the single letter index and then summing contributions from all words using the Plethystic exponential. Combining the contributions shown in Table~\ref{free}, we find that the index is
\be
I = \PE{ \frac{\left(p+p^{-1}\right) \sqrt{q_1 q_2 q_3}+q_1 q_2 q_3-\left(q_1 q_2+q_2 q_3+q_1 q_3\right)}{\left(1-q_1\right) \left(1-q_2\right) \left(1-q_3\right)} } \; . 
\label{(2,0)tensorindex}
\ee
Note that the denominator factors in the single letter index arise from summing up the action of holomorphic derivatives on the single letter contributions. 

On the other hand, the supersymmetric partition function of the tensor multiplet on $S^1 \times S^5$ is conjectured to be captured exactly by the partition function of 5d $\cN=2$ SYM on $S^5$ with gauge group $U(1)$. In order to relate the parameters appearing in the two partition functions, we define
\be
q_j = e^{-\beta \omega_j} \ ,\qquad p=e^{-\beta m} \, .
\ee
The parameter $\beta > 0$ is the radius of the circle $S^1$, which determines the 5d gauge coupling by the formula $g^2 = 2\pi \beta$. The parameters $\omega_j$ become squashing parameters for the metric on $S^5$ and $m$ is a real mass parameter for the adjoint $\cN=1$ hypermultiplet inside the $\cN=2$ tensor multiplet.

The $S^5$ partition function $Z$ was computed in~\cite{Kim:2012qf} using supersymmetric localization. The result was found to be proportional to the superconformal index $I$ given in equation~\eqref{(2,0)tensorindex} with a pre-factor that may be interpreted in terms of a supersymmetric Casimir energy. The result is \footnote{The notations here and in reference~\cite{Kim:2012qf} are related by $\omega_1 = 1+ a$, $\omega_2=1 + b$, $\omega_3 = 1 + c$ and $\delta^2 =\frac{1}{4}-m^2$. We have relaxed the relation $a+b+c=0$ imposed in \cite{Kim:2012qf}.}
\be
Z =  e^{- \beta E(1)} I \;,
\ee
where 
\be
E(1) = - \frac{1}{48 \, \omega_1\omega_2\omega_3} \Bigg[ \, \sigma_1^2\sigma_2^2 - \sum\limits_{i<j} \omega_i^2\omega_j^2 + \frac{1}{4} \bigg( \sum\limits_j \omega_j^2 - \sigma_1^2 - \sigma_2^2\bigg)^2 \, \Bigg] \;,
\label{6d-abelian-casimir}
\ee
is the supersymmetric Casimir energy. In writing this expression, we defined new chemical potentials $\sigma_1 \equiv \frac{1}{2}\sum_j \omega_j - m$ and $\sigma_2 \equiv \frac{1}{2}\sum_j \omega_j + m$, which are the chemical potentials conjugate to the R-symmetry generators $r_1$ and $r_2$ in the definition of the superconformal index. In other words, the superconformal index~\eqref{6dindexdef} is written as
$ \tr_{\cH_Q} (-1)^F e^{-\beta(\sum_j \omega_j h_j + \sum \sigma_a r_a ) }$
together with the constraint $\sigma_1+\sigma_2 = \sum_j \omega_j$. We use the notation $E(1)$ since this is the contribution to the supersymmetric Casimir energy from a single M5-brane.

Now let us compare the supersymmetric Casimir energy~\eqref{6d-abelian-casimir} with the anomaly polynomial of the free tensor multiplet (one M5-brane)~\cite{Witten:1996hc},
\be\label{eq:anomaly-I8}
A_{8}(1) = \frac{1}{48}\left[ p_2(NM) - p_2(TM) +\frac{1}{4}(p_1(NM) - p_1(TM) )^2 \right] \; .
\ee
In this expression, $TM$ and $NM$ denote respectively the tangent and normal bundles to the six-manifold $M$ where the brane is supported, and $p_j(V)$ is the $j$-th Pontryagin class of a real vector bundle $V$, which is a polynomial of degree $2j$. It is clear that the structure of the supersymmetric Casimir energy is mirrored in the anomaly polynomial.

To make the connection precise, we extend the anomaly polynomial~\eqref{eq:anomaly-I8} to an equivariant form on $\mathbb{R}^6$ with respect to the $U(1)^4$ action generated by the combinations of bosonic generators appearing in the superconformal index. There is a single fixed point at the origin of $\mathbb{R}^6$. Therefore, the equivariant integral of the anomaly polynomial can be computed using the fixed point theorem. This amounts to replacing the Chern roots of $TM$ with the chemical potentials $\omega_j$ and those of $NM$ with $\sigma_a$, and then dividing by the equivariant Euler class at the origin. Explicitly, we have
\begin{alignat}{3}
& p_1(NM) \longrightarrow \sigma_1^2+\sigma_2^2\;, \qquad\qquad
&& p_1(TM) \longrightarrow\sum_j \omega_j^2\;, \\
& p_2(NM) \longrightarrow \sigma_1^2\sigma_2^2\;,
&& p_2(TM) \longrightarrow \sum_{i<j} \omega_i^2 \omega_j^2 \, .
\end{alignat}
Making these replacements and dividing by the equivariant Euler class $e(TM)= \omega_1\omega_2\omega_3$, we find
\be
E(1) = -  \int A_8(1) \, , 
\ee
in agreement with our proposal (up to a conventional minus sign in the definition of the anomaly polynomial).

\subsubsection{Prediction for interacting theories}
\label{sec:predint}

Having confirmed our proposal for the free tensor multiplet, we can now make a prediction for the supersymmetric Casimir energy of the interacting 6d $\cN=(2,0)$ theories. The interacting theories are classified by a choice of simply-laced Lie algebra $\mathfrak{g}$.\footnote{One can of course also take direct sums of interacting theories and free tensor multiplets.} The group theoretic quantities associated to the simply-laced Lie algebras that we need in what follows are summarized in Table \ref{rhdg}.


\begin{table}[ht]
\centering
\begin{tabular}{|c|c|c|c|c|}
\hline
$\mathfrak{g}$ & $r_{\mathfrak{g}}$ & $d_{\mathfrak{g}}$ & $h^\vee_{\mathfrak{g}}$ & $\ell_i$ \\[0.2ex]
\hline
$A_{N-1}$ & $N-1$ & $N^2-1$ & $N$ & $2,3,\ldots,N$\\
$D_{N}$ & $N$ & $N(2N-1)$ & $2N-2$ & $2,4,\ldots,2N-2$ and $N$ \\
$E_6$ & $6$ & $78$ & $12$ & $2,5,6,8,9,12$ \\
$E_7$ & $7$ & $133$ & $18$ & $2,6,8,10,12,14,18$ \\
$E_8$ & $8$ & $248$ & $30$ & $2,8,12,14,18,20,24,30$ \\ [0.2ex]
\hline
\end{tabular}
\caption{The rank $r_\mathfrak{g}$, dimension $d_\mathfrak{g}$, dual Coxeter number $h^\vee_\mathfrak{g}$ and exponents $\{\ell_i \}_{i=1,\ldots,r_\mathfrak{g}} $ of the simply-laced Lie algebras.}
\label{rhdg}
\end{table}


The anomaly polynomial of the interacting theory is~\cite{Harvey:1998bx,Intriligator:2000eq,Yi:2001bz}
\be
A_8( \mathfrak{g} ) = r_{\mathfrak{g}} \, A_8(1) + d_{\mathfrak{g}} \,h^\vee_{\mathfrak{g}} \, \frac{p_2(NM)}{24}\;,
\ee
where $r_{\mathfrak{g}}$, $d_{\mathfrak{g}}$ and $h^\vee_{\mathfrak{g}}$ are the rank, dimension and dual Coxeter number of the simply-laced Lie algebra $\mathfrak{g}$, respectively. We should mention that, as far as we are aware, this formula for the anomaly polynomial is conjectural for the $E$-type theories. 

Performing the equivariant integral as explained above, we arrive at the conjecture that the supersymmetric Casimir energy of an interacting $(2,0)$ theory is
\bea
E( \mathfrak{g} ) & = - \int A_8( \mathfrak{g} ) = r_{\mathfrak{g}} \, E(1) - d_{\mathfrak{g}} \, h^\vee_{\mathfrak{g}} \, \frac{\sigma_1^2\sigma_2^2}{24 \, \omega_1\omega_2\omega_3} \,,
\label{nonabelian20-casimir}
\eea
where $E(1)$ is the supersymmetric Casimir energy of the Abelian tensor multiplet theory given in equation~\eqref{6d-abelian-casimir}.  

This prediction is very difficult to check because there is no Lagrangian construction in six dimensions that could be used to evaluate the partition function. Instead, we will use the conjecture that certain protected observable of the interacting 6d $\cN = (2,0)$ theories on a circle are captured by computations in 5d maximal SYM~\cite{Douglas:2010iu,Lambert:2010iw}. In particular, we suppose that the supersymmetric partition function on $S^1 \times S^5$ is equivalent to the partition function of 5d maximal SYM on $S^5$ with an appropriate identification of parameters. The latter can be computed by supersymmetric localization which reduces the path integral of the theory to a matrix integral~\cite{Kim:2012ava,Kim:2012qf,Minahan:2013jwa} (see also~\cite{Hosomichi:2012ek,Kallen:2012cs,Kallen:2012va,Imamura:2012xg,Imamura:2012bm} for related work). In practice, the resulting matrix integral cannot be evaluated explicitly for general values of the parameters, at least with present technology. In what follows, we will consider two simplifications of the problem that overcome this obstacle.

\subsubsection{Chiral algebra limit}
\label{sec:6dchiral}

We first consider a special limit of the superconformal index introduced in~\cite{Kim:2012qf,Kim:2013nva} where the matrix integral arising from localization of the $S^5$ path integral can be evaluated explicitly. This limit is
\be
p \to \sqrt{ q_1q_2 / q_3 } \;,
\ee
or equivalently 
\be
m \to \frac{1}{2}(\omega_1+\omega_2-\omega_3) \, .
\label{eq:limit}
\ee 
In this limit, the superconformal index and partition function preserve a second supercharge $Q_{++-}^{+-}$, which ensures additional cancellations in the matrix model and leads to a dramatic simplification of the result. This limit plays an important role in the ``chiral algebra" construction of~\cite{Beem:2014kka} and therefore we refer to it as the chiral algebra limit.

Let us first focus on the interacting theory of type $A_{N-1}$. The $S^1 \times S^5$ partition function is captured by the $S^5$ partition function of five-dimensional maximal SYM with gauge group $SU(N)$. In the limit~\eqref{eq:limit} the partition function reduces to the matrix integral
\be
\frac{1}{(\omega_1\omega_2)^{ \frac{N-1}{2} } } \int \frac{d^{N-1}a}{N!} \prod_{ i < j } \left[ 4 \sinh \frac{\pi}{\omega_1} (a_i-a_j) \, \sinh \frac{ \pi}{ \omega_2 }(a_i-a_j) \right]  
\frac{e^{- \frac{2\pi^2}{\beta \omega_1\omega_2\omega_3} \sum_i a_i^2}}{\eta\left( 2\pi \im / \beta\omega_3 \right)^{N-1}}\;,
\ee
where $\sum_i a_i = 0$ and $\eta(\tau)$ is the Dedekind eta function. Since the instanton contributions (the part of the integrand involving the Dedekind eta functions) are independent of $a_i$, the matrix integral is a sum of Gaussian integrals and can be evaluated explicitly. Remarkably, the result is proportional to a Plethystic exponential
\be
Z_{A_{N-1}} = q^{- c_{A_{N-1}} / 24} \, \mathrm{PE}\Bigg[ \, \frac{q^2+q^3+\cdots+q^N}{1-q} \, \Bigg] \;,
\label{eq:6dschurA}
\ee
where
\be
c_{A_{N-1}} =  (N-1)+ N(N^2-1) \frac{(\omega_1+\omega_2)^2}{\omega_1\omega_2} \, .
\ee

For a general simply-laced Lie algebra, the $S^1 \times S^5$ partition function is expected to be given the following generalization of equation~\eqref{eq:6dschurA}
\be
Z = q^{- c_{\mathfrak{g}} / 24} \, \mathrm{PE}\Bigg[ \, \frac{1}{1-q} \sum_{i=1}^{r} q^{\ell_i} \, \Bigg] \;,
\label{eq:6dschurgen}
\ee
where
\be
c_{\mathfrak{g}} =  r_{\mathfrak{g}} + d_{\mathfrak{g}}  \, h_{\mathfrak{g}}^\vee \, \frac{(\omega_1+\omega_2)^2}{\omega_1\omega_2}\;,
\label{eq:walgebrac}
\ee
and $\{ \ell_i \}$ are the exponents shown inTable~\ref{rhdg}. This formula can be checked by explicit computation which can be performed for the theories of type $A_{N-1}$ and $D_{N}$. The result is conjectural for the $E$-type theories since the instanton contributions are unknown. This expression is the vacuum character of the $\cW$-algebra of type $\mathfrak{g}$ with central charge $c_\mathfrak{g}$ found in~\cite{Drukker:2010jp}. In the limit $\omega_1 = \omega_2=1$, corresponding to a round five-sphere, this result can be interpreted in terms of the ``chiral algebra" construction~\cite{Beem:2014kka}.

The supersymmetric Casimir energy extracted from the partition function~\eqref{eq:6dschurgen} is thus
\be
E(\mathfrak{g}) =   - \frac{\omega_3}{24} c_{\mathfrak{g}}\;.
\label{eq:6dchiralE}
\ee
It is simple to check that our prediction~\eqref{nonabelian20-casimir} for the general supersymmetric Casimir reduces to this formula in the limit~\eqref{eq:limit}. It is also interesting to note that the supersymmetric Casimir energy of the six-dimensional theory \eqref{eq:6dchiralE} is proportional to the usual non-supersymmetric Casimir energy of a two-dimensional Toda CFT of type $\mathfrak{g}$ with central charge $\eqref{eq:walgebrac}$.

\subsubsection{General parameters}
\label{sec:6dlargeN}

The 6d supersymmetric Casimir energy can be extracted from the $S^1 \times S^5$ partition function in the limit that the radius of $S^1$ becomes large, $\beta\to\infty$. Therefore it is not necessary to compute the full partition function in order to extract the supersymmetric Casimir energy. In this section, we will attempt to compute the supersymmetric Casimir energy with general parameters turned on by focusing on the $\beta \to \infty$ limit.

We will focus exclusively on the 6d theory of type $\mathfrak{g}=A_{N-1}$. We will assume that the $S^1 \times S^5$ partition function is captured exactly by the partition function of 5d $SU(N)$ maximal SYM theory on $S^5$ with gauge coupling 
\be
g^2 = 2\pi \beta\;,
\ee
and for convenience, we set the radius of $S^5$ to 1. Then the supersymmetric Casimir energy in 6d is identified with the strong coupling limit of the free energy in 5d. In 5d terminology, the statement is
\be
\log Z_{S^5} \to -  \frac{g^2}{2\pi} E + \cdots \qquad \mathrm{as} \qquad g^2 \to \infty \, .
\ee
To compute the leading behavior at strong coupling, we will first include only the classical and 1-loop contributions to the partition function, for a moment forgetting the contributions from instantons saddle points. Later, we will argue that instantons give a certain correction to the free energy by comparing it with the conjectured free energy and with its special limit considered in the previous subsection.

Similar computations have been performed before in the literature for the large $N$ free energy of the $S^5$ partition function~\cite{Kallen:2012zn,Jafferis:2012iv,Minahan:2013jwa}.
These references considered the parameter regime where the instanton contributions are suppressed, and thus the partition function becomes a simple matrix integral involving only classical and 1-loop contributions. We will compare our result with their free energy and see a perfect agreement at large $N$. In particular, the instanton corrections to the free energy in our result begin to appear at order $\mathcal{O}(N)$, which is subleading in the large $N$ expansion. This is therefore consistent with the expectation that the instanton contributions are suppressed at large $N$.

The exact partition function $Z_{S^5}$ can be computed using the technique of supersymmetric localization~\cite{Kallen:2012cs,Kallen:2012va,Kim:2012ava,Imamura:2012bm,Kim:2012qf}.
The path integral localizes to constant vacuum expectation values for the scalar field $\langle \phi \rangle = a$ in the $\mathcal{N}=1$ vector multiplet. In addition, there are singular instanton saddle points localized at the three fixed circles of the Killing vector $\sum_{j=1}^3\omega_j \, h_j$ generated by $Q^2$.
As described above, we will first omit the instanton contributions. The full perturbative partition function takes the form~\cite{Imamura:2012bm,Kim:2012qf,Lockhart:2012vp}
\be
	Z_{S^5}(m,\vec\omega,\beta)=\frac{1}{(\omega_1\omega_2)^{ \frac{N-1}{2} } } \int \frac{d^{N-1}a}{N!} e^{-\frac{2\pi^2}{\beta \omega_1\omega_2\omega_3}(a,a)} \prod_{i=1}^3Z_{\rm 1-loop}^{(i)}(a,m,\vec\omega) \ .
\label{eq:ZS51loop}	
\ee
The integration is over the scalar vev $a$ in the Cartan subalgebra of $SU(N)$ (in our conventions $a$ is real) and $(\, , \, )$ denotes the inner product on the Cartan subalgebra normalized such that the norm of all simple coroots is 2.

The 1-loop contributions factorize into three fixed point contributions $Z_{\rm 1-loop}^{(i)}$ where $i$ labels one of three fixed points on the base of the Hopf fibration $S^5 \to \mathbb{CP}^2$. 
Collecting the three 1-loop determinants, we obtain
\bea\label{eq:1-loop-determinant}
  \hspace{-0.5cm}\prod_{i=1}^3Z^{(i)}  
   = \left( \frac{\lim_{x\rightarrow 0}S_3(x)/x}{S_3(\tilde{m})} \right)^{N-1} \prod_{i>j}^N
  \frac{S_3\big(\im a_{ij}|\vec{\omega}\big)S_3\big(-\im a_{ij}|\vec{\omega}\big)}
  {S_3\big(\tilde{m}+\im a_{ij}|\vec{\omega}\big)S_3\big(\tilde{m} - \im a_{ij}|\vec{\omega}\big)} \ ,
\eea
where $\im\equiv \sqrt{-1}$, $a_{ij}\equiv a_i-a_j$, and $\tilde{m} \equiv m +\frac{\omega_1+\omega_2+\omega_3}{2}$.  Here $S_3(z|\vec\omega)$ is the triple-sine function whose definition and properties we summarize in Appendix~\ref{app:2}.

We will now evaluate the integral \eqref{eq:ZS51loop} in the strong coupling limit, $\beta \rightarrow \infty$. 
If we assume that the vector multiplet scalar vev $a$ is very large while other parameters remain of order one, we can approximate the triple sine function as
\bea
{\rm log} S_3(\im a|\vec\omega)  
  & \stackrel{{\rm sgn}(a)= \pm1}{\approx} -\frac{\pi}{6\omega_1\omega_2\omega_3}\bigg(|a|^3\pm \im\frac{3}{2}(\omega_1+\omega_2+\omega_3)|a|^2 \\
  &\qquad\qquad-\frac{1}{2}(\omega_1^2+\omega_2^2+\omega_3^2+3\omega_1\omega_2+3\omega_2\omega_3+3\omega_3\omega_1)|a|  \\
  & \qquad \qquad  \qquad \qquad \mp \frac{\im}{4}(\omega_1+\omega_2+\omega_3)(\omega_1\omega_2+\omega_2\omega_3+\omega_3\omega_1) \bigg) \ .
\eea
If we further restrict the scalar $a_i$ to a Weyl chamber where $a_i > a_j$ for $i>j$, then the perturbative partition function can be approximated as
\be
	Z_{S^5} = \int [da] \ e^{-\frac{2\pi}{\omega_1\omega_2\omega_3}f(a,\vec\omega,m)} \ ,
\ee
where
\begin{equation}
\begin{split}
f(a,\vec\omega,m) &\approx  
 \frac{\pi}{\beta}\sum_{i=1}^Na_i^2 + \frac{1}{12}\sum_{i>j}\left(2(a_{ij})^3-(a_{ij}+\im\tilde{m})^3-(a_{ij}-\im\tilde{m})^3\right) \\[5pt]
 & -\frac{\tilde{m}(\omega_1+\omega_2+\omega_3)}{2}\sum_{i>j}a_{ij} +\mathcal{O}(\beta^0) 
 =  \frac{\pi}{\beta} \sum_{i=1}^Na_i^2 -\frac{\sigma_1\sigma_2}{2}\sum_{i>j}a_{ij} +\mathcal{O}(\beta^0) \ .
\end{split}
\end{equation}
One can evaluate this partition function using the saddle point approximation.
Note that the saddle point solution exists only when $\sigma_1\sigma_2 >0$  since the scalar $a_i$ are already ordered. Assuming $\sigma_1\sigma_2 > 0$, we find the solution
\be
	a_j = \frac{\beta\sigma_1\sigma_2}{4\pi}(2j-N-1) \ ,
\ee
which is consistent with our assumption of large $a_j$ at large $\beta$.
Plugging this into the partition function, we finally obtain
\be
-{\rm log} Z_{S^5} = -\beta\frac{(N^2-1)N \sigma_1^2\sigma_2^2}{24\,\omega_1\omega_2\omega_3} +\mathcal{O}(\beta^0)\;,
\ee
and hence
\be
E_{\mathrm{pert}}(A_{N-1}) = - \frac{(N^2-1)N \sigma_1^2\sigma_2^2}{24\,\omega_1\omega_2\omega_3} \, .
\label{eq:Epert}
\ee
We emphasize that this is the result for the supersymmetric Casimir energy we obtain by removing the instanton contributions to the 5d partition function.

We can now compare this result with our conjecture for the general supersymmetric Casimir energy~\eqref{nonabelian20-casimir}. For type $\mathfrak{g}=A_{N-1}$ equation~\eqref{nonabelian20-casimir} reduces to
\be
E(A_{N-1}) = (N-1)E(1) -\frac{(N^2-1)N\sigma_1^2\sigma_2^2}{24\, \omega_1\omega_2\omega_3} \;.
\label{eq:EANm1}
\ee
Clearly, we find agreement between our perturbative result \eqref{eq:Epert} and the second term in the right hand side of \eqref{eq:EANm1}. The first term is $(N-1)$ copies of the supersymmetric Casimir energy of a free tensor multiplet. It is tempting to conjecture that this is the contributions from instantons. More generally, we can conjecture the instantons in the 5d computation to contribute $r_{\mathfrak{g}} E(1)$ to the supersymmetric Casimir energy $E(\mathfrak{g})$. Although we could not perform a complete calculation including instantons, we view the harmony between the general formula in~\eqref{nonabelian20-casimir} and the perturbative result in \eqref{eq:Epert} as strong evidence in favor of our conjecture.

Finally, we mention that our result is consistent with the large $N$ free energy computed in~\cite{Minahan:2013jwa,Kallen:2012zn}. The instanton corrections are indeed suppressed at large $N$, appearing at $\cO(N)$ compared to the leading perturbative contribution at $\cO(N^3)$.
We also find that the conjectured instanton correction, i.e. the first term on the right hand side of \eqref{eq:EANm1}, is consistent with the exact result in the special limit~\eqref{eq:limit}. In this limit, the instanton correction to the free energy becomes
\be\label{eq:instanton-correction}
r_{\mathfrak{g}} E(1) ~ 
\longrightarrow ~ -r_{\mathfrak{g}}\frac{\omega_3}{24}\,.
\ee
In the previous section, we saw that the instanton contribution in the special limit simplifies to
$\eta(2\pi \im/\beta\omega_3)^{r_{\mathfrak{g}}}$. After performing the modular transformation, one can easily check that the exact instanton correction to the free energy in the limit $\beta \to \infty$ is precisely the formula~\eqref{eq:instanton-correction}.
Furthermore, it also agrees with the exact instanton correction of the abelian $U(1)$ 5d SYM at strong coupling, computed in~\cite{Kim:2012qf}.

\subsection{$\cN=(1,0)$ supersymmetry}
\label{sec:(1,0)6d}

The 6d $\cN=(1,0)$ superconformal algebra is $\mathfrak{osp}(8^*|2)$ with bosonic subalgebra $\mathfrak{so}(2,6)\oplus \mathfrak{usp}(2)$. We denote the conformal generators as above and $r$ is the Cartan generator of the $\mathfrak{usp}(2)=\mathfrak{su}(2)$ R-symmetry. There are eight chiral Poincar\'e supercharges in the two-dimensional representation of $\mathfrak{usp}(2)$, which we denote by $Q_{h_1,h_2,h_3}^r$ with $h_1h_2h_3<0$ and $r=\pm\frac{1}{2}$. In addition, there are eight conformal supercharges with the opposite chirality, $h_1h_2h_3>0$.

We will define the 6d $\cN=(1,0)$ superconformal index using the supercharge $Q\equiv Q^+_{---}$. This generates the subalgebra
\be
	\{Q,Q^\dagger\} = \Delta -4r -(h_1+h_2+h_3) \, .
\ee
There are three Cartan generators $h_j+r$ commuting with this supercharge and we will introduce fugacities $q_j$ for them. The superconformal index is defined as
\be
I = {\rm Tr}_{\cH_Q}(-1)^F \prod_{j=1}^3 q_j^{h_j+r}\, z^{f} \, .
\ee
Unlike $\cN = (2,0)$ supersymmetry, $\cN = (1,0)$ superconformal theories can have non-trivial global (non-R) symmetries. The exponent $f$ above stands for the Cartan generators of the global symmetry algebra and $z$ is the corresponding fugacity. 

\subsubsection{$E$-string Theories}
\label{sec:estring}

A large class of 6d $\mathcal{N}=(1,0)$ SCFTs have been argued to exist using F-theory constructions~\cite{Heckman:2015bfa} as well as constraints from anomaly cancellations~\cite{Bhardwaj:2015xxa}. Here, we focus exclusively on a simple class known as `$E$-string' theories. In M-theory, they appear on the worldvolume of $N$ coincident M5-branes embedded in an end-of-the world brane with $E_8$ symmetry. 

As the transverse space is $\mathbb{R}^4\times \mathbb{R}_{>0}$ we expect an internal symmetry $\mathfrak{so}(4) \simeq \mathfrak{su}(2)_1\times \mathfrak{su}(2)_2$ rotating the $\mathbb{R}^4$ directions. We identify the first factor $\mathfrak{su}(2)_1$ with the $\mathfrak{usp}(2)$ R-symmetry in the superconformal algebra, while $\mathfrak{su}(2)_2$ becomes an additional global symmetry. The $E$-string theories also correspond to small $E_8$ instantons in $E_8\times E_8$ heterotic string theory and are expected to have an $E_8$ global symmetry~\cite{Witten:1995gx,Ganor:1996mu,Seiberg:1996vs}.

The anomaly polynomials of $E$-string theories have been computed in~\cite{Ohmori:2014pca} (see also~\cite{Ohmori:2014kda} for more general $\cN = (1,0)$ theories).
Expanding in powers of $N$, the anomaly polynomial takes the form\footnote{The tensor multiplet anomaly polynomial $A_8(1)$ from equation (\ref{eq:anomaly-I8}) and $I_8$ in reference~\cite{Ohmori:2014pca} are related by $A_8(1) = -I_8 + \frac{p_2(NM)}{24}$.}
\be\label{eq:anomaly-E-string}
	A_{E_8+{\rm free}}(N) =\frac{N^3}{6}p_2(NM) + \frac{N^2}{2} e(NM)A_4  + N\left(\frac{A_4^2}{2} - \frac{p_2(NM)}{24} + A_8(1) \right) \ ,
\ee
where $A_8(1)$ is the anomaly polynomial of a free $\cN=(2,0)$ tensor multiplet \eqref{eq:anomaly-I8}, $e(NM)$ is the Euler class of the normal bundle, and $A_4 \equiv \frac{1}{4}\left(p_1(NM)+p_1(TM) + {\rm Tr} F^2\right)$.
The two-form $F$ is the background curvature for the $E_8$ global symmetry.
The subscript ``free" implies that it involves the free hypermultiplet contribution. 

We now compute the equivariant integral of this anomaly polynomial. We can recycle computations involving $TM$ and $NM$ from the previous section, by the replacement
\be
	\sigma_1 = \frac{1}{2}\sum_{j=1}^3\omega_j - \mu \ , \qquad \sigma_2 = \frac{1}{2}\sum_{j=1}^3\omega_j +\mu \, ,
\ee
where $\mu$ is the chemical potential for the $\mathfrak{su}(2)_2$ global symmetry and $\sum_j\omega_j$ is the chemical potential for the R-symmetry $\mathfrak{su}(2)_1$. In addition, we have chemical potentials $m_1,\ldots,m_8$ for the $E_8$ global symmetry. The equivariant integral of the anomaly polynomial on $\mathbb{R}^6$ is
\be
\begin{aligned}
\int A_{E_8+{\rm free}}(N) 
& = \frac{N^3  \sigma _1^2 \sigma _2^2}{6\,\omega_1\omega_2\omega_3}  -\frac{N^2 \sigma _1 \sigma _2 }{8 \, \omega_1\omega_2\omega_3}  \Bigg[\sigma _1^2+\sigma _2^2+ \sum_j \omega _j^2 + 2 \sum_a m_a^2 \Bigg] \\
& + \frac{N}{\omega_1\omega_2\omega_3 } \left[ \frac{1}{32} \left( \sigma _1^2+\sigma _2^2 +\sum_j \omega _j^2+2 \sum_a m_a^2 \right)^2 - \frac{\sigma_1^2\sigma_2^2}{24} \right. \\
& + \left. \frac{1}{4} \left( \sigma _1^2+\sigma _2^2-\sum_j \omega _j^2\right)^2+\sigma _1^2 \sigma _2^2-\sum_{i<j} \omega _i^2 \omega _j^2  \right] \, .
\end{aligned} 
\label{eq:6d(1,0)pred}
\ee

The $E$-string theories do not have a Lagrangian construction in 6d. However, upon circle compactification, it is believed that they have a low-energy description in terms of 5d $\mathcal{N}=1$ SYM with $Sp(2N)$ gauge group, an antisymmetric hypermultiplet, and $N_f=8$ fundamental hypermultiplets~\cite{Ganor:1996pc,Seiberg:1996bd}. The non-trivial Wilson line along the compactified circle breaks the UV $E_8$ global symmetry to $SO(16)$ symmetry in 5d. It is expected that the full $E_8$ global symmetry is restored in the UV limit of the 5d gauge theory by strong coupling dynamics involving non-perturbative effects.

We are not aware of a limit analogous to the one in Section \ref{sec:6dchiral} for the E-string SCFTs and thus we proceed as in Section \ref{sec:6dlargeN} and compute the free energy of the 5d theory on a squashed $S^5$ in the strong coupling limit and compare it with the anomaly polynomial. As in Section \ref{sec:6dlargeN}, we first compute the free energy contribution only from the perturbative partition function and later make a conjecture for the instanton correction.
The perturbative partition function takes the following matrix integral expression:
\be
	Z_{S^5}^{E_8}(m_a,\vec\omega,\beta) = \int [da] e^{-\frac{4\pi^3r}{g^2\omega_1\omega_2\omega_3}(a,a)} \times \frac{\prod_{e \in {\rm root}} S_3\left(\im(e,a)|\vec{\omega}\right)' }
	{ \prod_{a=1}^{8+1}\prod_{\rho\in R_a} S_3\left(\tilde{m}_a+\im(\rho,a) | \vec{\omega}\right) } \ ,
\ee
where the primed function is defined for zero modes such as $S_3(0)'\equiv \lim_{x\rightarrow0} S_3(x)/x$. This theory has 8 fundamental hypermultiplets with mass $m_1,\ldots,m_8$ and an antisymmetric tensor hypermultiplet with mass $m_9\equiv\mu$. We have defined shifted masses $\tilde{m}_a \equiv m_a+\frac{\omega_1+\omega_2+\omega_3}{2}$. $R_a$ stands for representations of the hypermultiplets.

We can evaluate the matrix integral in the strong coupling limit $g\rightarrow \infty$. If we assume again that the scalar $a$ takes a large saddle point expectation value, then the integral reduces to
\bea
	Z_{S^5}^{E_8} &= \int [da] \ e^{-\frac{4\pi^3}{g^2\omega_1\omega_2\omega_3}f(a,\vec\omega,m_a)} \ , \\ 
	f(a,\vec\omega,m_a) &\equiv \frac{4\pi^2}{g^2}\sum_{i=1}^N a_i^2 +f_V(a) + f_{\rm anti}(a,\mu) + \sum_{b=1}^{8} f_{\rm fund}(a,m_b) \ ,
\eea
where,
\bea
	f_V(a) &\equiv \frac{1}{6}\sum_{i>j}^N\bigg[|a_i\pm a_j|^3-\frac{\mathcal{E}}{2}|a_i\pm a_j|\bigg] + \frac{1}{6}\sum_{i=1}^N\left[|2a_i|^3 -\frac{\mathcal{E}}{2}|2a_i|\right] +\mathcal{O}(g^0) \ , \\
	f_{\rm anti}(a,\mu) &\equiv -\frac{1}{6}\sum_{i> j}^N\left[|a_i\pm a_j|^3-3\left[\mu^2-\frac{1}{4}(\sum_{k=1}^3\omega_k)^2\right]|a_i\pm a_j|-\frac{\mathcal{E}}{2}|a_i\pm a_j|\right] +\mathcal{O}(g^0)\ ,  \\
	f_{\rm fund}(a,m_b) &\equiv -\frac{1}{6}\sum_{i=1}^N\left[|a_i|^3 -3\left[m_b^2-\frac{1}{4}(\sum_{k=1}^3\omega_k)^2\right]|a_i| -\frac{\mathcal{E}}{2}|a_i|\right] +\mathcal{O}(g^0)\;,
\eea
are the contributions from the vector multiplet, the antisymmetric hypermultiplet, and the fundamental hypermultiplets, respectively. To simplify the expression, we have defined 
\be
\mathcal{E}\equiv \sum_{i=1}^3\omega_i^2 +3 \sum_{i>j}\omega_i\omega_j\, .
\ee
We have also used the shorthand notation: $|a\pm b|^n \equiv |a+b|^n + |a-b|^n$.  One can easily see that the cubic terms cancel, while the remaining terms reduce to
\begin{multline}
	f(a,\vec\omega,m_a) \\
	= \frac{4\pi^2}{g^2} \sum_{i=1}^Na_i^2 - \frac{1}{2}\sigma_1\sigma_2\sum_{i>j}^N|a_i\pm a_j| - \frac{1}{2}\sum_{i=1}^N \sum_{b=1}^8\left[\frac{1}{4}(\sum_{k=1}^3\omega_k)^2-m_b^2\right]|a_i| + \frac{\mathcal{E}}{2}\sum_{i=1}^N | a_i| +\mathcal{O}(g^0)\;.
\end{multline}
We now choose a Weyl chamber in which $a_i>a_j$ for $i>j$ and $a_i>0$.
The solution of the saddle point equation is 
\be
	a_i = \frac{g^2}{16\pi^2}\left[2\sigma_1\sigma_2(i-1) - \sum_{b=1}^8m_b^2 + \sum_{j=1}^3\omega_j^2 + \sum_{j>k}^3\omega_j\omega_k \right] \ .
\ee
This solution makes sense only when all masses are much smaller than the $\omega_j$'s.

Inserting this solution back into the partition function, we find the free energy of the E-string theory when $g^2\rightarrow \infty$
\bea
- {\rm log} Z^{E_8}_{S^5} & = 
- \frac{N^3g^2\sigma_1^2\sigma_2^2}{24\pi \omega_1\omega_2\omega_3} 
- \frac{N^2g^2\sigma_1\sigma_2}{32\pi\omega_1\omega_2\omega_3}\left[\sigma_1^2+\sigma_2^2 + \sum_{j=1}^3\omega_j^2 + 2\sum_{b=1}^8m_b^2\right] \\
& \hspace{1.5cm} -\frac{Ng^2}{96\pi \omega_1\omega_2\omega_3}\left[\frac{3}{4} \left[\sigma_1^2+\sigma_2^2 + \sum_{j=1}^3\omega_j^2 + 2\sum_{b=1}^8m_b^2\right]^2 - \sigma_1^2\sigma_2^2\right] +\mathcal{O}(g^0)\ .
\eea
We now identify the 5d gauge coupling with the radius of the 6d circle by $g^2 = 4\pi \beta$.
Note that this differs by a factor 2 from the relation in the $\cN = (2,0)$ case. With this identification, the perturbative contribution to the supersymmetric Casimir energy is
\bea
	E^{E_8}_{\mathrm{pert}} & = - \frac{N^3 \sigma_1^2\sigma_2^2}{6 \, \omega_1\omega_2\omega_3} - \frac{N^2\sigma_1\sigma_2}{8 \, \omega_1\omega_2\omega_3}\left[\sigma_1^2+\sigma_2^2 + \sum_{j=1}^3\omega_j^2 + 2\sum_{b=1}^8m_b^2\right] \\
	& \hspace{1.5cm} -\frac{N}{24 \, \omega_1\omega_2\omega_3}\left[ \frac{3}{4} \left[\sigma_1^2+\sigma_2^2 + \sum_{j=1}^3\omega_j^2 + 2\sum_{b=1}^8m_b^2\right]^2 - \sigma_1^2\sigma_2^2\right] +\mathcal{O}(g^0)\ .
\eea

A comparison with the equivariant integral~\eqref{eq:6d(1,0)pred} shows that 
\be
	E^{E_8}_{\rm pert}(N) -  N \int A_8(1) = -  \int A_{E_8+{\rm free}}(N)  \, .
\ee
Therefore we find agreement of our perturbative computation with the prediction for the full supersymmetric Casimir energy of the $E$-string theory up to a correction $N \int A_8(1)$, which is $-N$ times the contribution from a free tensor multiplet~(\ref{6d-abelian-casimir}). We view this as strong evidence in favor of our prediction. As in Section \ref{sec:6dlargeN}, full consistency requires that the correction
\be
	E_{\rm inst}^{E_8} = - N \int A_8(1)\;,
\ee
is the contribution to the supersymmetric Casimir energy from instantons.

\section{Four dimensions}
\label{sec:4d}

\subsection{$\cN=1$ supersymmetry}
\label{sec:4dN=1Lag}

The four-dimensional $\cN=1$ superconformal algebra is $\mathfrak{su}(2,2|1)$, which has a maximal bosonic subalgebra $\mathfrak{su}(2,2) \oplus \mathfrak{u}(1)$. We will denote the Cartan generators of the conformal subalgebra $\mathfrak{su}(2,2)$ by $(\Delta,h_1,h_2)$, where $\Delta$ is the dilatation generator and $(h_1,h_2)$ generate rotations in two orthogonal planes. The $\mathfrak{u}(1)$ R-symmetry generator is $r$.

We define the $\mathcal{N}=1$ superconformal index using the supercharge with quantum numbers $h_1 = h_2 = - \frac{1}{2}$ and $r=1$. This supercharge generates the subalgebra 
\be
\{Q,Q^\dagger\} = \Delta - h_1 - h_2 - \frac{3}{2}r\;,
\ee
and the Cartan generators commuting with the supercharges $Q$ and $Q^\dagger$ are $h_1+\frac{r}{2}$ and $h_2+\frac{r}{2}$, together with the Cartan generators $f$ of any flavor symmetry. The superconformal index is defined by
\be\label{N=1index}
	\cI = {\rm Tr}_{\cH_Q}(-1)^{\rm F} p^{h_{1} + \frac{r}{2}}q^{h_{2}+\frac{r}{2}} a^{f} \, ,
\ee
where $\cH_Q$ is the subspace of states in radial quantization that saturate the unitarity bound $\Delta -h_1-h_2-\frac{3}{2}r \geq 0$. We have introduced fugacities $p$, $q$ and $a$ respectively for the Cartan generators $h_1+\frac{r}{2}$, $h_2+\frac{r}{2}$ and $f$. For convergence we assume that $|p|, |q| < 1$.

\subsubsection{Lagrangian theories}

For an $\cN=1$ SCFT that has a weakly-coupled Lagrangian description in the UV, the superconformal index can be computed by enumerating gauge invariant operators in the UV and then identifying the correct IR R-symmetry. 

Let us consider a theory with a compact semi-simple gauge group $G$, flavor symmetry $F$, and chiral multiplets transforming in a complex representation $\cR$ of $G \times F$. We introduce an additional fugacity $\zeta$ valued in the maximal torus $T_G \subset G$. The superconformal index is then a matrix integral 
\be
\label{4dN=1int}
I = \int \, [d\zeta] \cdot \hat{\Delta}(\zeta) \cdot \mathcal{I}^{\rm vm}(\zeta) \cdot  \mathcal{I}^{\rm cm}(\zeta) \;,
\ee
where
\be
\hat{\Delta}(\zeta) \equiv  \frac{1}{| W |} \prod_{e \in \hat{\Delta}^+} (1-\zeta^e)(1-\zeta^{-e})
\label{eq:Haar}
\ee
is the Haar measure on $G$. The notation $\hat{\Delta}^+$ denotes the set of positive roots and $| W|$ is the dimension of the Weyl group.

The integrand in \eqref{4dN=1int} consists of contributions from vector multiplets and chiral multiplets, which may be computed as Plethystic exponentials of the single-letter indices. The contributions are
\bea
\label{4dN=1vm}
\mathcal{I}^{\rm vm}  
& = \mathrm{PE}\left[ \frac{2pq-p-q}{(1-p)(1-q)} \chi_{\rm adj}(\zeta)  \right]\;, \\
\mathcal{I}^{\rm cm} 
& = \mathrm{PE}\left[ \sum_{(\rho,\rho') \in \cR } \frac{(p\,  q)^{\frac{r_{\rho,\rho'}}{2}} \zeta^\rho  a^{\rho'} - (pq)^{1-\frac{r_{\rho,\rho'}}{2}} \zeta^{-\rho} a^{-\rho'} }{(1-p)(1-q)}    \right] \, ,  
\eea
where $\chi_\mathrm{adj}(\zeta)$ is the character of the adjoint representation of $G$ and $(\rho,\rho')$ are the weights of the representation $\cR$ of $G \times F$. $r_{\rho,\rho'}$ is the $\mathfrak{u}(1)$ charge of the chiral multiplet at the IR fixed point, which can be determined in a given theory by anomaly cancellation and/or a-maximization~\cite{Intriligator:2003jj}.

 The partition function of a Lagrangian $\cN=1$ theory on $S^1 \times S^3$ may also be computed using supersymmetric localization~\cite{Assel:2014paa}. The parameters of the $S^1 \times S^3$ partition function are related to the parameters of the superconformal index by 
\be
\label{eq:4dN=1rep}
p=e^{-\beta\omega_1}\;, \qquad q=e^{-\beta\omega_2}\;, \qquad a=e^{-\beta m}\;,
\ee
where $\beta>0$ is the radius of $S^1$, $(\omega_1,\omega_2)$ are squashing parameters for the geometry of $S^3$, and $m$ are expectation values of background vector multiplets for flavor symmetries. Similar to the superconformal index, the path integral on $S^1\times S^3$ reduces to a matrix integral
\be
\label{eq:N=1-index-integral}
Z = \int \, [d\zeta] \cdot \hat{\Delta}(\zeta) \cdot Z^{\rm vm}(\zeta) \cdot  Z^{\rm cm}(\zeta) \;,
\ee
where $\zeta_a = e^{-\beta z_a}$ with $z_{a} \sim z_{a} + 2\pi \im/\beta$ is the gauge holonomy around $S^1$.
The integrand is a product of 1-loop determinants from the vector multiplets and chiral multiplets, which take the form of infinite products over KK-momenta around $S^1$ and require careful regularization. In reference~\cite{Assel:2015nca} (see also \cite{Ardehali:2015hya}), a $\zeta$-function regularization scheme compatible with the supercharge $Q$ used in localization was proposed and we will employ this regularization scheme in what follows.

 The regularized 1-loop determinants for the vector multiplets and chiral multiplets take the form
\be
\label{eq:4d-vector-1loop}
Z^{\rm vm} = e^{- \beta E^{\rm vm} } I^{\rm vm}\;, \qquad 
Z^{\rm cm} = e^{- \beta E^{\rm cm} } I^{\rm cm} \;,
\ee
where $I^{\rm vm}$ and $I^{\rm cm}$ are the contributions to the superconformal index given in~\eqref{4dN=1vm}. As shown in reference~\cite{Assel:2015nca}, the functions appearing in the exponentials are\footnote{In reference~\cite{Assel:2014paa} there were additional contributions in the exponentials at order $\cO(\beta^{-1})$, which are absent in the regularization scheme introduced in~\cite{Assel:2015nca}}
\bea
\label{4dN=1casimir}
E^{\rm vm} & = \sum_{e\in \Delta} f\left( \la z,e\ra +\frac{\omega_1+\omega_2}{2} \right) \, ,\\
E^{\rm cm} & = \sum_{(\rho,\rho')\in \cR} f\left( \la z,\rho\ra + \la m,\rho'\ra + (r_{\rho,\rho'}-1)\frac{\omega_1+\omega_2}{2} \right) \, ,
\eea
where
\be
f(u) = \frac{u^3}{6 \, \omega_1\omega_2} - \frac{\omega_1^2+\omega_2^2}{24 \, \omega_1\omega_2}u \;,
\label{eq:fdef}
\ee
and $\la \, ,\, \ra$ denotes the canonical pairing between a Cartan subalgebra and its dual. In a consistent theory, there are no cubic or mixed 't Hooft anomalies for the gauge symmetry $G$, meaning that the total contribution $E = E^{\rm vm}+E^{\rm cm}$ is independent of the gauge chemical potential $z$. The prefactor $e^{-\beta E}$ can then can be pulled outside the matrix integral and the $S^1 \times S^3$ partition function is directly proportional to the superconformal index, $Z = e^{- \beta E} I$. The function $E$ is the supersymmetric Casimir energy on $S^1 \times S^3$. 

We shall now identify $E$ with the equivariant integral of the anomaly polynomial of the corresponding $\cN=1$ SCFT. In four dimensions, anomalies arise from massless chiral fermions coupled to background gauge fields. For a chiral fermion in a representation ${\cR}$ of the group $K$, the six-form anomaly polynomial is
\be
A_6 = \left[\hat{A}(TM)\cdot {\rm Tr} ( e^{F} )\right]_6 = \frac{{\rm Tr} (F^3)}{6} - \frac{p_1(TM)}{24} {\rm Tr} (F)  \ ,
\label{eq:N=16form}
\ee
where $\hat{A}(TM)$ is the A-roof genus of a four-dimensional manifold $M$, $p_1(TM)$ is the first Pontryagin class, and $F$ is the curvature of the associated $K$-bundle corresponding to the representation $\cR$. The subscript $|_6$ means we extract the six-form component in the polynomial expansion in the curvatures.

We consider $M = \mathbb{R}^4$ and work equivariantly with respect to $K \times U(1)^2$ where $U(1)^2$ are the rotations generated by $(h_1,h_2)$. We introduce equivariant parameters $m$ for $K$ and $(\omega_1,\omega_2)$ for $U(1)^2$ and evaluate the equivariant integral using the fixed point theorem. There is a single fixed point at the origin of $\mathbb{R}^4$. Therefore, the equivariant integral amounts to replacing the Chern roots of the characteristic classes by the corresponding equivariant parameters, and dividing by the equivariant Euler class at the origin, $e(TM) = \omega_1 \, \omega_2$. For the characteristic classes appearing in~\eqref{eq:N=16form} we have 
\be
p_1(TM) \longrightarrow \omega_1^2 +\omega_2^2 \qquad \tr(F^n) \longrightarrow \sum_{\rho \in \cR} \, \la m , \rho \ra^n \, \, .
\ee
Therefore the equivariant integral of the anomaly polynomial is
\be
\int A_6 = \sum_{\rho \in \cR} \, \left[ \, \frac{\la m,\rho \ra^3}{6\,\omega_1\omega_2} - \frac{\omega_1^2+\omega_2^2}{24\,\omega_1\omega_2} \, \la m,\rho \ra \, \right]  = \sum_{\rho \in \cR} f\big(  \la m , \rho \ra \big)
\ee
where the function $f(u)$ is defined in~\eqref{eq:fdef}.

\medskip

\begin{table}[h]
\centering
\begin{tabular}{ | c | c | c | c | c |} \hline
  \quad $U(1)_1$ \quad & \quad $U(1)_2$ \quad & \quad $U(1)_r$ \quad & \ $F$ \quad & \ $G$ \quad  \\ \hline
  $\omega_1$ \quad & $\omega_2$ & $\frac{\omega_1+\omega_2}{2}$ & $m$ & $z$  \\ \hline
\end{tabular}
\caption{Equivariant parameters in the 4d $\cN=1$ superconformal index.}
\label{N=1eq}
\end{table}

Now we consider the case relevant for the $\cN=1$ superconformal index where we take the $K$-bundle to be a product of the gauge group $G$, a global symmetry group $F$, and the R-symmetry $U(1)_r$, $K = G \times F \times U(1)_r$. The corresponding equivariant parameters are summarized in Table~\ref{N=1eq}. The contributions from fermions in vector and chiral multiplets are:
\begin{itemize}
\item A vector multiplet contains a chiral fermion in the adjoint representation of $G$ with $U(1)_r$ charge $1$. 
\item A chiral multiplet whose lowest component has $U(1)_r$ charge $r$ contains a chiral fermion with charge $r-1$.
\end{itemize}
Summing these contributions to the anomaly polynomial, we find that its equivariant integral is
\bea
\label{4dN=1casimirA}
\int A_6 
& = \sum_{e\in \Delta} f\left( \la z,e\ra +\frac{\omega_1+\omega_2}{2} \right) 
& + \sum_{(\rho,\rho')\in \cR} f\left( \la z,\rho\ra + \la m,\rho'\ra + (r_{\rho,\rho'}-1)\frac{\omega_1+\omega_2}{2} \right) \, .
\eea
This is exactly the supersymmetric Casimir energy $E$, i.e. the sum of the two terms in \eqref{4dN=1casimir}. We therefore conclude that for $\cN=1$ SCFTs realized by Lagrangian theories in the UV, the supersymmetric Casimir energy is an equivariant integral of the anomaly polynomial.

Note that the anomaly polynomial encodes potential contributions from cubic and mixed 't Hooft gauge anomalies, as well as global anomalies. If they were present, $E$ would contain terms cubic or quadratic in the gauge holonomy $z$, which would violate the periodicity $z_{a} \sim z_{a} + 2\pi \im/\beta$ and imply that the holonomy integral in the $S^1 \times S^3$ partition function is ill defined. This is consistent with the fact that the superconformal index computation for a theory with broken gauge or R-symmetry does not make sense. For a consistent theory, $E$ is independent of $z$.

\subsubsection{Example: $\cN=1$ superconformal QCD}
\label{sec:N=1SQCD}

Before writing a general expression for the supersymmetric Casimir energy, we consider a concrete example. Let us consider $\mathcal{N}=1$ SQCD with $ G = SU(N_c)$ gauge group and $F = SU(N_f)_1\times SU(N_f)_2\times U(1)_B$ flavor symmetry. The theory has $N_f$ chiral multiplets $Q$ in the fundamental representation and $N_f$ chiral multiplets $\tilde Q$ in the anti-fundamental representation of $SU(N_c)$. The quarks $Q$ and $\tilde{Q}$ have $+1$ and $-1$ baryon charge respectively, and R-charge $r=(N_f-N_c)/N_f$. 

To simplify our expressions, we find it convenient to introduce the notation 
\be
\sigma = \frac{1}{2}(\omega_1+\omega_2)\;,
\ee
for the chemical potential conjugate to $U(1)_r$. With this notation, the supersymmetric Casimir energy, or equivalently the equivariant integral of the anomaly polynomial, is given by
\begin{eqnarray}
	E &=& \sum_{i\neq j}^{N_c} \left[\frac{(z_i-z_j+\sigma)^3}{6\, \omega_1\omega_2} - \frac{\omega_1^2+\omega_2^2}{24\, \omega_1\omega_2}(z_i-z_j+\sigma) \right] + \frac{(N_c-1)\sigma}{12}\\
	&& + \sum_{i=1}^{N_c}\sum_{j=1}^{N_f} \left[\frac{(z_i+m_{j}+b+(r-1)\sigma)^3}{6 \, \omega_1\omega_2} - \frac{\omega_1^2+\omega_2^2}{24 \, \omega_1\omega_2}(z_i+m_{j}+b+(r-1)\sigma) \right] \nonumber \\
	&& + \sum_{i=1}^{N_c}\sum_{j=1}^{N_f} \left[\frac{(\tilde{m}_{j}-z_i-b+(r-1)\sigma)^3}{6 \, \omega_1\omega_2} - \frac{\omega_1^2+\omega_2^2}{24 \, \omega_1\omega_2}(\tilde{m}_j-z_i-b+(r-1)\sigma) \right]\;, \nonumber
\end{eqnarray}
where $m_i$ and $\tilde{m}_j$ (subject to $\sum_i m_i=\sum_j\tilde{m}_j=0$) are the chemical potentials for the flavor symmetries $SU(N_f)_1$ and $SU(N_f)_2$ and $b$ is the chemical potential for $U(1)_B$.

Let us now expand this formula and identify the contributions from the various anomalies that can occur. It is straightforward to show that
\begin{eqnarray}
\label{eq:om1om2E}
	\omega_1\omega_2 \, E &=& \left((r-1)N_f+N_c\right) \sigma \sum_{i=1}^Nz_i^2 + \frac{k_{111}}{6} \sum_{i=1}^{N_f}m_i^3+ \frac{k_{222}}{6} \sum_{i=1}^{N_f}\tilde{m}_i^3 \nonumber\\
	 && +k_{11r} \, \sigma \sum_{i=1}^{N_f}m_i^2 + k_{22r} \, \sigma \sum_{i=1}^{N_f}\tilde{m}_i^2
	 + k_{11B}\, b \sum_{i=1}^{N_f}m_i^2 + k_{22B}\, b \sum_{i=1}^{N_f}\tilde{m}_i^2+ \frac{k_{BBr}}{2} \sigma \, b^2 \nonumber\\
	 && + \frac{k_{rrr}}{6} \sigma^3  - \frac{k_r}{24}(\omega_1^2+\omega_2^2)\sigma \ ,
\end{eqnarray}
where
\begin{eqnarray}
	&&k_{111}=k_{222}= N_c \,, \qquad 
	k_{11r}=k_{22r} = (r-1)\frac{N_c}{2} \,, \nonumber\\ 
	&&k_{11B}=k_{22B} = \frac{N_c}{2} \,, \qquad k_{BBr} = 2(r-1)N_fN_c\;, \\
	&& k_{rrr} = 2(r-1)^3N_fN_c + N_c^2 - 1 \,, \qquad
	k_r = 2(r-1)N_fN_c + N_c^2 -1 \ , \nonumber
\end{eqnarray}
are the cubic and linear 't Hooft anomaly coefficients for currents labeled by the corresponding subscript, i.e. $k_{11B}$  is the cubic anomaly coefficient from a triangle diagram with two $SU(N_f)_1$ and one $U(1)_B$ currents. The first term on the right hand side of \eqref{eq:om1om2E} is quadratic in $z$ and corresponds to the quadratic gauge anomaly from the $SU(N_c)^2 \times U(1)_r$ triangle diagram. Indeed, this term vanishes with the correct R-charge assignment $r=(N_f-N_c)/N_f$. The remaining terms on the right-hand side of \eqref{eq:om1om2E} encode all non-vanishing global anomalies for this theory. Each anomaly is described by a triangle diagram with a current at each vertex. The coefficient $k_r$ corresponds to the triangle diagram involving a $U(1)_r$ current and two energy momentum tensors.

\subsubsection{General formula}
\label{sec:gen4d}

Suppose that we have a 4d $\mathcal{N}=1$ SCFT with $U(1)_R$ superconformal R-symmetry and global symmetry $F =\prod_a F_a \times \prod_I U(1)_I$ where  $U(1)_I$ are Abelian flavor symmetries, and $\prod_a F_a$ is a semi-simple flavor symmetry. Expanding the general expression \eqref{eq:N=16form}, we find that the supersymmetric Casimir energy is
\begin{align}
\label{eq:gen4dN1E}
E= \int A_6 =& \ \frac{k_{rrr}}{6\omega_1\omega_2}\sigma^3 +\frac{k_{rrI}}{2\omega_1\omega_2} \sigma^2m_I+\frac{k_{rIJ}}{2\omega_1\omega_2} \sigma m_Im_J +\frac{k_{IJK}}{6\omega_1\omega_2} m_Im_Jm_K  \cr
&\ + \frac{k_{rab}}{2\omega_1\omega_2} \sigma \la  m_a, m_b \ra + \frac{k_{Iab}}{2\omega_1\omega_2} m_I \la m_a , m_b \ra \cr
&\ - \frac{k_r}{24\omega_1\omega_2} \sigma(\omega_1^2+\omega_2^2)- \frac{k_I}{24\omega_1\omega_2} m_I(\omega_1^2+\omega_2^2) \; ,
\end{align}
where $k_{ABC}$ and $k_A$ are the cubic and linear 't Hooft anomalies. When the theory has a Lagrangian description one has $k_{ABC}=\text{Tr}_f(ABC)$ and $k_{A}=\text{Tr}_f(A)$ where the trace is over the chiral fermions $f$ in the theory. Notice however that the anomaly polynomial is also applicable and useful for interacting theories without a known Lagrangian description. Note that if the flavor symmetry contains $SU(N)$ factors, there may be additional cubic anomaly terms which we have omitted from~\eqref{eq:gen4dN1E}.

Note that the relation between the conformal and 't Hooft anomalies in a 4d $\cN=1$ theory is
\begin{equation}\label{ack}
a = \frac{9}{32} k_{rrr} - \frac{3}{32} k_r\;, \qquad c = \frac{9}{32} k_{rrr} - \frac{5}{32}k_r\;.
\end{equation}
In the absence of flavor symmetries, or after setting the chemical potentials for any flavor symmetries to zero, one can use the relation \eqref{ack} to reproduce the following result for the supersymmetric Casimir energy~\footnote{This result also agrees with the SUSY Casimir energy in~\cite{Assel:2014paa}, up to $\mathcal{O}(\beta^{-1})$ terms.}
\begin{equation}
\begin{split}
E =  \frac{2}{3}(a-c)(\omega_1+\omega_2)  + \frac{2}{27}(3c-2a) \frac{(\omega_1+\omega_2)^3}{\omega_1\omega_2} \; ,
\end{split}
\end{equation}
which was derived in reference~\cite{Assel:2015nca,Ardehali:2015hya}.

\subsection{$\cN=2$ supersymmetry}
\label{sec:4dN=2}

The 4d $\cN=2$ superconformal algebra is $\mathfrak{su}(2,2|2)$, which has the maximal bosonic subalgebra $\mathfrak{su}(2,2) \oplus \mathfrak{su}(2)_R \oplus \mathfrak{u}(1)_r$. The Cartan generators of the conformal algebra $\mathfrak{su}(2,2)$ are denoted as in the previous section, while the R-symmetry generator in the Cartan of $\mathfrak{su}(2)_R$ is denoted by $R$ and the superconformal R-symmetry $\mathfrak{u}(1)_r$ by $r$.

We will define the superconformal index using the supercharge $Q$ with quantum numbers $h_1=h_2=-\frac{1}{2}$, $R=\frac{1}{2}$ and $r=-\frac{1}{2}$. This supercharge generates the commutator
\be
\{ Q,Q^\dagger \} = \Delta - h_1-h_2 - 2R+r\;, 
\ee
and a linearly independent basis of Cartan generators commuting with $Q$ are $h_1-r$, $h_2-r$ and $r+R$, together with the generators $f$ of any flavor symmetry. The superconformal index is defined as
\be
I = \tr_{\cH_Q}(-1)^F p^{h_1-r}q^{h_2-r}t^{r+R} a^f \, ,
\ee
where $\cH_Q$ is the subspace of states in radial quantization that saturate the bound $\Delta - h_1-h_2- 2R+r \geq 0$. We have introduced fugacities $p$, $q$, $t$ and $a$ for the Cartan generators commuting with $Q$. For convergence we assume that $|p|,|q|,|t|,|pq/t|<1$. 

\subsubsection{Lagrangian theories}
\label{sec:4dN=2lagrangian}

In this section, we will focus on 4d $\cN=2$ SCFTs that have UV Lagrangian descriptions constructed from $\mathcal{N}=2$ vector multiplets and hypermultiplets. We consider a theory with semi-simple gauge group $G$, flavor symmetry $F$, and hypermultiplets in a complex representation $\cR$ of $F \times G$. For simplicity, we will not consider the possibility of half-hypermultiplets.

Introducing an additional fugacity $\zeta$ valued in the maximal torus $T_G \subset G$, the superconformal index can be expressed as a matrix integral
\be
I = \int \, [d\zeta] \cdot \hat{\Delta}(\zeta) \cdot \cI^{\rm vm}(\zeta) \cdot  \cI^{\rm hm}(\zeta) \;,
\label{4dN=2int}
\ee
where the Haar measure was defined in equation~\eqref{eq:Haar}. The contributions to the integrand from vector multiplets and hypermultiplets are
\bea
\cI^{\rm vm} & = {\rm PE} \left[ \left(-\frac{p}{1-p}-\frac{q}{1-q} + \frac{pq/t-t}{(1-p)(1-q)} \right) \chi_{\rm adj}(\zeta) \right]\;, \\
\cI^{\rm hm} & = {\rm PE} \left[  \frac{\sqrt{t}-pq/\sqrt{t}}{(1-p)(1-q)}  \sum_{(\rho,\rho') \in \cR} ( \zeta^\rho a^{\rho'} + \zeta^{-\rho} a^{-\rho'}  )\right]  \, ,
\eea
where $\chi_{\mathrm{adj}}(\zeta)$ is the character of the adjoint representation of the gauge group $G$ and $(\rho,\rho')$ are the weights of the representation $\cR$.

We now compare the superconformal index with the $S^1\times S^3$ partition function. To make the connection, we introduce chemical potentials
\be
\label{eq:4dN=2rep}
p=e^{-\beta\omega_1}\;, \qquad q=e^{-\beta\omega_2}\;, \qquad t = e^{-\beta \gamma}\;, \qquad a=e^{-\beta m}\;.
\ee
It is also convenient to define $\sigma = \gamma - \sum_j \omega_j$ so that the superconformal index becomes
\be
I = \tr_{\cH_Q}(-1)^F e^{-\beta( \sum_j \omega_j h_j  + \gamma R + \sigma r + mf) }\, .
\ee
In the $S^1 \times S^3$ partition function, $\omega_j$ becomes squashing parameters, $m$ are expectation values for background flavor vector multiplets, and $\gamma, \sigma$ are the background expectation values of background R-symmetry vector multiplets.

The partition function of a Lagrangian $\cN=2$ theory can be computed by viewing it as an $\cN=1$ theory with distinguished flavor symmetries due to the extra R-symmetry. The contributions to the integrand from the 1-loop determinants of vector multiplets and hypermultiplets are
\be
Z^{\rm vm} = e^{- \beta E^{\rm vm}} \mathcal{I}^{\rm vm}\;, \qquad\qquad
Z^{\rm hm} = e^{- \beta E^{\rm hm}} \mathcal{I}^{\rm hm}\;,
\ee
where
\bea
E^{\rm vm} & = - \sigma  \left[ \; \sum_{e \in \Delta^+} \la e,z \ra^2  + \frac{n_V}{12} ( \gamma ^2+\gamma  \sigma +\omega _1 \omega _2 ) \; \right]\;, \\
E^{\rm hm} & = \sigma \Bigg[ \; \frac{1}{2} \sum_{(\rho,\rho')\in \cR}  \big(\la \rho,z\ra+\la\rho',m\ra \big)^2 + \frac{n_H}{24}( \sigma^2 - \omega_1^2-\omega_2^2) \; \Bigg] \, ,
\eea
where $n_V = \mathrm{dim}(G)$ is the number of vector multiplets and $n_H = \mathrm{dim}(\cR)$ is the number of hypermultiplets. It is again illuminating to express the exponential contributions in terms of the function $f(z)$ defined in equation~\eqref{eq:fdef}. We find that
\bea
E^{\rm vm} 
& = \sum_{\lambda \in \rm{adj} } \!\bigg[  f\left( \, \la \lambda , z \ra - \frac{\sigma}{2} +\frac{\gamma }{2} \, \right)
+ f\left( \, \la  \lambda , z \ra - \frac{\sigma}{2}  - \frac{\gamma }{2} \, \right)  \bigg] \; , \\
E^{\rm hm}  & = \sum_{(\rho,\rho')\in \cR} \!\bigg[ f\left( \, \la \rho,z\ra + \la \rho',m \ra + \frac{\sigma}{2} \, \right)
+ f\left( \, -\la\rho,z\ra - \la \rho',m\ra + \frac{\sigma}{2} \, \right) \bigg] \, .
\label{eq:4dN=2Es}
\eea

It is straightforward to identify the terms in~\eqref{eq:4dN=2Es} with the contributions from the fermions in the hypermultiplets and the vector multiplets to the equivariant integral of the anomaly polynomial. The contribution to the equivariant integral of the anomaly polynomial from a single fermion in a 4d $\cN=2$ supermultiplet is
\be
f\left( \, \la \rho , z \ra +  \la \rho' ,m \ra + r \left(\gamma -\omega _1-\omega _2\right) + R \gamma \, \right)\;,
\ee
where $\rho$ is the gauge weight, $\rho'$ the flavor weight, and $(R,r)$ are the R-symmetry charges of the fermion. The contributions from vector multiplets and hypermultiplets are as follows:
\begin{itemize} 
\item From the vector multiplet, we have a pair of chiral fermions with $(R,r) = (\pm\frac{1}{2},-\frac{1}{2})$ for each weight $\lambda$ of the adjoint representation.
\item  From the hypermultiplet, we have a pair of conjugate fermions with $(R,r)=(0,\frac{1}{2})$ for each weight $(\rho,\rho')$ of the complex representation $\mathcal{R}$.
\end{itemize}

Summing the contributions from all fermions $\psi$, the supersymmetric Casimir energy can be written
\be
E =  \frac{\sigma\gamma^2}{\omega_1\omega_2} \frac{ \mathrm{Tr}_{\psi}(r R^2)}{2}  + \frac{\sigma^3 }{\omega_1\omega_2}\frac{ \mathrm{Tr}_{\psi}(r^3)}{6} - \frac{\sigma(\omega_1^2+\omega_2^2) }{\omega_1\omega_2}\frac{ \mathrm{Tr}_{\psi}(r)}{24}
+ \frac{\sigma}{\omega_1\omega_2} \sum_{\psi} \frac{r_{\psi} \la \rho_{\psi}, m \ra^2}{2} \, .
\ee
This formula can be expressed in terms of the representation $\cR$ of $G \times F$ as follows
\bea
E = & -\frac{1}{8} \mathrm{dim}(\cR)  \frac{\sigma(\sigma+\omega_1+\omega_2)^2}{\omega_1\omega_2}  + \frac{1}{24}(\mathrm{dim}(\cR)- \mathrm{dim}(G) )\frac{\sigma(\sigma^2-\omega_1^2-\omega_2^2) }{\omega_1\omega_2} \\
& +  \frac{\sigma}{2 \, \omega_1\omega_2} \sum_b k_{rbb} \la m_b , m_b \ra +  \frac{\sigma}{2 \, \omega_1\omega_2} \sum_{I,J} k_{rIJ}  \, m_I m_J  \, ,
\label{4dN=2lag}
\eea
where, in order to express the flavor symmetry anomalies, we have unpackaged the flavor symmetry as a product of simple and Abelian factors $F =  \prod_b F_b  \, \times \, \prod_IU(1)_I$. The numbers $k_b$ and $k_{IJ}$ are the 't Hooft anomaly coefficients for the triangle diagrams $U(1)_r \times F_b^2$ and $U(1)_r \times U(1)_I \times U(1)_J$ respectively. Explicitly, we have
\begin{itemize}
\item 
The $U(1)_r \times F_b^2$ anomaly is
\be
k_{rbb} =  \sum_j T(\cR_j^{(b)})\, ,
\label{eq:ksimple}
\ee
where we decompose $\cR \to \oplus_j \cR_j^{(b)}$ into irreducible representations of the simple factor $F_b$, and $T(\cR_j^{(b)})$ is the index of the representation normalized so that the index of the adjoint representation is the dual Coxeter number $h^\vee$. 
\item The $U(1)_r \times U(1)_I \times U(1)_J$ anomaly is
\be
k_{rIJ} =  \sum_j q_j^{(I)} q_j^{(J)}\, ,
\label{eq:kabelian}
\ee
where the summation $j$ is over hypermultiplets and $q_j^{(I)}$ is the charge of the $j$-th hypermultiplet under $U(1)_I$.
\end{itemize}

\subsubsection{General formula}
\label{sec:4dN=2gen}

Based on the Lagrangian computations, or the generic form of the anomaly polynomial with 4d $\cN=2$ superconformal symmetry, we can now make the following prediction for the supersymmetric Casimir energy of a general 4d $\cN=2$ SCFT,
\bea
E =&  \frac{1}{2}(c-2a) \frac{\sigma(\sigma+\omega_1+\omega_2)^2}{\omega_1\omega_2}  + (c-a) \frac{\sigma(\sigma^2-\omega_1^2-\omega_2^2) }{\omega_1\omega_2} \\
& +
 \frac{\sigma}{4 \, \omega_1\omega_2} \sum_b k_{rbb} \la m_b , m_b \ra +  \frac{\sigma}{4 \, \omega_1\omega_2} \sum_{I,J} k_{rIJ} \, m_I m_J \, ,
\label{4dN=2pred}
\eea
where, as above, the summation $b$ is over simple factors and $I$ is over Abelian factors of the flavor symmetry group. The anomaly coefficients $a$, $c$, $k_{rbb}$ and $k_{rIJ}$ are defined directly in the conformal field theory in terms of correlation functions of the R-symmetry and flavor symmetry currents.

In a Lagrangian theory,
\bea
c-a &  = \frac{1}{24}(\mathrm{dim}(\cR)- \mathrm{dim}(G) ) \;,
 \\
c - 2a & = -\frac{1}{4} \mathrm{dim}(\cR)\;,
\eea
and $k_{rbb}$ and $k_{rIJ}$ are defined in equations~\eqref{eq:ksimple} and~\eqref{eq:kabelian} respectively, in which case we reproduce~\eqref{4dN=2lag}.

\subsubsection{Example: $\cN=2$ superconformal QCD}
\label{sec:N=2SQCD}

As an illustration of a Lagrangian theory, we briefly consider $\mathcal{N}=2$ superconformal QCD, that is, $SU(N)$ gauge theory with $2N$ fundamental hypermultiplets. This theory arises in class $\cS$ from a sphere with two maximal and two minimal punctures and has flavor symmetry (at least) $SU(N) \times SU(N) \times U(1) \times U(1)$. We introduce corresponding chemical potentials $y_i$, $z_i$, $b_1$ and $b_2$. 

The supersymmetric Casimir energy is found to be
\bea
E(N) =  -\frac{N^2-1}{8} \frac{\sigma(\sigma+\omega_1+\omega_2)^2}{\omega_1\omega_2} + \frac{N^2+1}{24}  \frac{\sigma(\sigma^2-\omega_1^2-\omega_2^2) }{\omega_1\omega_2} \\  + \frac{N}{2} \frac{\sigma}{\omega_1\omega_2}\sum_{i=1}^N(y_i^2+z_i^2) + \frac{N^2}{2} \frac{\sigma}{\omega_1\omega_2}(b_1^2+b_2^2)\, ,
\eea
which agrees with~\eqref{4dN=2pred} since
\be
c = \frac{1}{6}(2N^2-1)\;, \qquad a = \frac{1}{24}(7N^2-5)\;,
\qquad
k_{SU(N)} = N\;, \qquad k_{U(1)} = N^2\, .
\ee
This agreement was of course guaranteed by the general construction of Section~\ref{sec:4dN=2lagrangian}. A much more non-trivial check would be to compute the supersymmetric Casimir energy of a theory without a known Lagrangian construction.

\subsubsection{Example: $\cT_3$}
\label{sec:T3}

We now want to test our conjecture for the supersymmetric Casimir energy with a ``non-Lagrangian" example. We consider the $\cT_3$ theory with $E_6$ flavor symmetry discovered by Minahan and Nemeschansky~\cite{Minahan:1996fg}. This theory arises in class $\cS$ by compactifying the 6d $\cN=(2,0)$ theory of type $A_2$ on a sphere with three maximal punctures~\cite{Gaiotto:2009we}. The flavor symmetry manifest in this construction is $SU(3)^3 \subset E_6$.

\begin{figure}[htp]
\centering
\includegraphics[width=4.5in]{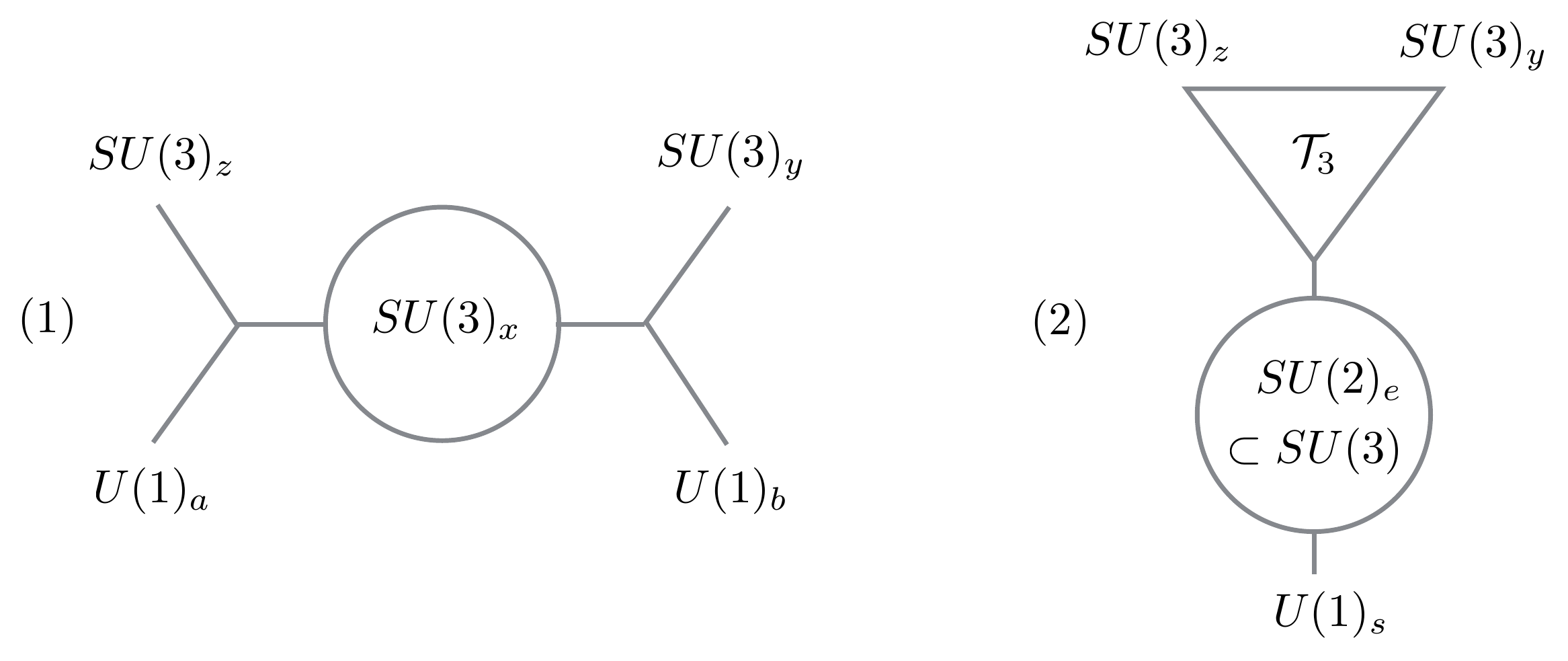}
\caption{S-duality transformation relating $SU(3)$ superconformal SQCD to a $\cT_3$ coupled to a fundamental $SU(2)$ hypermultiplet. Subscripts correspond to chemical potentials in main text.}
\label{fig:E6duality}
\end{figure}

The superconformal index of $\cT_3$ has been computed by exploiting consistency with $S$-duality in reference~\cite{Gadde:2010te}. The same idea can be used to compute the supersymmetric Casimir energy. In duality frame (1) we have $SU(3)$ superconformal SQCD. In duality frame (2) we have a fundamental hypermultiplet of $SU(2)$ coupled to $\cT_3$ by gauging an $SU(2) \subset SU(3)$ at one puncture. This is illustrated in Figure~\ref{fig:E6duality}.

We introduce chemical potentials $a$ and $b$ for the $U(1)$ symmetries at the two minimal punctures and $z_j$ and $y_j$ for the $SU(3)$ symmetries at the two maximal punctures. In duality frame (1), we further introduce chemical potentials $x_j$ for the $SU(3)$ gauge symmetry. The supersymmetric Casimir energy is
\bea
E_{(1)} & = \sum _{i=1}^3 \sum _{j=1}^3  f\left[ a+z_i-x_j + \frac{\sigma}{2}  \right] +f\left[ -a-z_i+x_j + \frac{\sigma}{2} \right] \\
& + \sum _{i=1}^3 \sum _{j=1}^3  f\left[ b+y_i+x_j + \frac{\sigma}{2} \right]  + f\left[ - b -y_i -x_j +\frac{\sigma}{2} \right] \\
& +\sum _{i=1}^3 \sum _{j=1}^3 f\left[  x_i-x_j + \frac{\sigma}{2}+\frac{\gamma }{2}  \right] + f\left[ x_i-x_j + \frac{\sigma}{2} - \frac{\gamma }{2}  \right] \\
& - f\left[ \frac{\sigma}{2}+\frac{\gamma }{2}  \right] - f\left[  \frac{\sigma}{2} - \frac{\gamma }{2}  \right] \, .
\eea
As a consistency check, it is straightforward to see that this expression is independent of $x_1$, $x_2$ and $x_3$ (here it is important that $x_1+x_2+x_3=0$). 

In duality frame (2), we introduce the chemical potential $e$ for the $SU(2) \subset SU(3)$ being gauged and a chemical potential $s$ for the $U(1)$ symmetry of the hypermultiplet. The supersymmetric Casimir energy in this frame is
\bea
E_{(2)} & = f\left[ e + s + \frac{\sigma}{2} \right] +f\left[ e - s + \frac{\sigma}{2} \right]
 + f\left[ - e + s + \frac{\sigma}{2} \right] +f\left[  -e - s + \frac{\sigma}{2} \right] \\
& +  f\left[ 2e+ \frac{\sigma}{2}+\frac{\gamma }{2}  \right] + f\left[ 2e + \frac{\sigma}{2} - \frac{\gamma }{2}  \right] 
+ f\left[ - 2e+ \frac{\sigma}{2}+\frac{\gamma }{2}  \right] + f\left[ - 2e + \frac{\sigma}{2} - \frac{\gamma }{2}  \right]\\
& + f\left[  \frac{\sigma}{2}+\frac{\gamma }{2}  \right] + f\left[  \frac{\sigma}{2} - \frac{\gamma }{2}  \right] + E_{\cT_3} \, ,
\eea
where $E_{\cT_3}$ is the supersymmetric Casimir energy of $\cT_3$.

We now want to compute $E_{\cT_3}$ by setting $E_{(1)} = E_{(2)}$. To compare the expressions, we note that the non-manifest $SU(3)$ chemical potentials of the $\cT_3$ theory are given by $\{w_1,w_2,w_3\}=\{r+e,r-e,-2r\}$ where $r = -\frac{1}{2}(a+b)$. Furthermore, we have $s = \frac{3}{2}(a-b)$. With these identifications, we find
\bea
\label{eq:EMN}
E_{\cT_3} & = \frac{3}{2} \frac{\sigma}{\omega_1\omega_2} \sum _{i =1}^3 \left(w_i^2+y_i^2+ z_j^2\right)  -\frac{5}{8} \frac{\sigma  \left(\sigma +\omega _1+\omega _2\right)^2 }{\omega_1\omega_2}+ \frac{11}{24} \frac{\sigma  \left(\sigma ^2-\omega _1^2-\omega _2^2\right)}{\omega_1\omega_2}\;.
\eea
Note that the dependence of the flavor parameters is
\be
\sum _{i < j} \left(w_i^2+y_i^2+z_i^2\right) =  \la m , m \ra^2\;,
\ee
where $m$ is the chemical potential for the $E_6$ flavor symmetry. The expression in \eqref{eq:EMN} is in precise agreement with the conjecture~\eqref{4dN=2pred} using the known conformal anomalies $c = \frac{13}{6}$ and $a = \frac{41}{24}$, and flavor anomaly $k_{E_6} = 3$.

\subsubsection{Schur Limit and Chiral Algebras}
\label{sec:schur}

Finally, we consider a particularly simple limit of the 4d $\cN=2$ superconformal index in order to make contact with the work~\cite{Beem:2013sza} on chiral algebras. This limit may be reached from our general construction by setting $q=t$. The combinations of Cartan generators appearing in the definition of the superconformal index now commute with an additional supercharge, leading to dramatic simplifications. In particular, the superconformal index depends only on $q$.

It was shown in~\cite{Beem:2013sza} that the superconformal index becomes the character of the vacuum representation $V_0$ of a 2d chiral algebra,
\be
I(q) = \mathrm{Tr}_{V_0} \,\left(  q^{L_0} \right)\, ,
\ee
whose 2d central charge is related to the 4d conformal anomaly by
\be
c_{\mathrm{2d}} = - 12 c\, .
\ee

Let us now consider the same limit of the supersymmetric Casimir energy, by setting $\gamma = \omega_2$. From the general formula~\eqref{4dN=2pred}, we find that the supersymmetric Casimir energy now depends only on $c$ (we turn off chemical potentials for flavor symmetries in this section) and in particular
\be
e^{-\beta E} = q^{c/2}\, .
\ee
Combining with the superconformal index, we find that the $S^1 \times S^3$ partition function is
\be
Z =  \mathrm{Tr}_{V_0} \, \left( q^{L_0 - c_{2d} / 24} \right)\, ,
\ee
which reproduces the character of the vacuum representation, but now including the conformal anomaly prefactor that is necessary for good modular properties. This may be a hint towards interesting ``modular" properties of the full $S^1 \times S^3$ partition function of 4d $\cN=2$ theories with general fugacities.

\subsection{$\mathcal{N}=4$ supersymmetry}
\label{sec:4dN=4}

As a final example in four dimensions, we consider $\mathcal{N}=4$ SYM with gauge group $G$. This theory has $\mathfrak{su}(2,2|4)$ superconformal algebra whose  bosonic subalgebra is $\mathfrak{so}(2,4)\times \mathfrak{so}(6)_R$. In this section, we denote the Cartan generators of the R-symmetry as $(R_1,R_2,R_3)$.

We will define a superconformal index with a supercharge $Q^{R_1R_2R_3}_{h_1h_2} = Q^{---}_{--}$ giving the commutator $\{Q,Q^\dagger\} = \Delta -h_1 -h_2 + R_1+R_2+R_3$. The superconformal index will count the protected states commuting with this supercharge.

The superconformal index is defined as
\be
	I^{\mathcal{N}=4} = {\rm Tr}(-1)^F e^{-\beta\omega_1h_1-\beta\omega_2h_2 -\beta m_1 R_1-\beta m_2 R_2-\beta m_3 R_3} \;,
\ee
where $\omega_{1,2},m_{1,2,3}$ are the chemical potentials for the four Cartan generators commuting with $Q$ and thus they are subject to the constraint $\omega_1+\omega_2+m_1+m_2+m_3=0$.

The $S^1\times S^3$ partition function of the $\mathcal{N}=4$ theory is computed in~\cite{Kinney:2005ej} from the UV free theory Lagrangian using a localization argument. Taking into account the regularization factors carefully, the partition function can be written as 
\be
	Z^{\mathcal{N}=4} = e^{-\beta E^{\mathcal{N}=4}} I^{\mathcal{N}=4} \ ,
\ee
where
\be
\label{eq:ENeq4}
	E^{\mathcal{N}=4}= d_G \frac{m_1m_2m_3}{2\omega_1\omega_2} \ .
\ee

The supersymmetric Casimir energy is again equivalent to the equivariant integral of the anomaly polynomial. The $\mathcal{N}=4$ vector multiplet contains 4 chiral fermions carrying the following R-charges:
\begin{center}
\begin{tabular}{|c|ccc|}
\hline
& $R_1$ & $R_2$ & $R_3$ \\
\hline
$\lambda$ & $\frac{1}{2}$ & $\frac{1}{2}$ & $\frac{1}{2}$ \\
$\chi_1$ & $-\frac{1}{2}$ & $-\frac{1}{2}$ & $\frac{1}{2}$ \\
$\chi_2$ & $\frac{1}{2}$ & $-\frac{1}{2}$ & $-\frac{1}{2}$ \\
$\chi_3$ & $-\frac{1}{2}$ & $\frac{1}{2}$ & $-\frac{1}{2}$ \\
\hline
\end{tabular}
\end{center}
The equivariant integral of the anomaly polynomial can be easily performed with these R-charge data, and one obtains
\be
	\int A_6^{\mathcal{N}=4}= \frac{d_G}{\omega_1\omega_2}\sum_{i=1}^4\left[\frac{\mu_i^3}{6} -\frac{(\omega_1^2+\omega_2^2)\mu_i}{24}\right] = d_G \frac{m_1m_2m_3}{2\omega_1\omega_2} \ ,
\ee
where $\mu_i$ are weights of the spinor representation of $SO(6)$ R-symmetry, i.e. $\mu_1=\frac{m_1+m_2+m_3}{2},\,\mu_2=\frac{-m_1-m_2+m_3}{2},\,\mu_3=\frac{m_1-m_2-m_3}{2},\,\mu_4=\frac{-m_1+m_2-m_3}{2}$. Indeed, this result agrees with the supersymmetric Casimir energy in \eqref{eq:ENeq4}.

\section{Two dimensions}
\label{sec:2d}

\subsection{$\cN=(0,2)$ supersymmetry}
\label{sec:elgen}

We consider the superconformal index (or ``flavored" elliptic genus) of 2d $\mathcal{N}=(0,2)$ SCFTs. At the end of the day, we want to compute the superconformal index in the ``NS sector". In this case, we define the superconformal index with respect to the supercharge $Q$ (sometimes also denoted as $G^-_{-\frac{1}{2}}$ in the super-Virasoro algebra, see for example \cite{Schwimmer:1986mf}) in radial quantization, which satisfies the algebra $[2\bar{L}_0,Q]=[R,Q]=Q$ and
\be
	\{Q,Q^\dagger\} = 2\bar{L}_0 - R \ ,
\ee
where $2\bar{L}_0 = \Delta-J$ is a combination of the scaling dimension $\Delta$ and the angular momentum $J$, and $R$ is the $U(1)_R$ R-charge. 

The 2d $\cN=(0,2)$ superconformal algebra has a one-parameter family of automorphisms, parametrized by an element $e^{2\pi i \eta} \in \mathbb{C}^*$, where $\eta$ is conventionally known as the ``spectral flow parameter". The corresponding one-parameter family of generators are 
\be
	\bar{L}_0^\eta = \bar{L}_0 + \left(\eta-\frac{1}{2}\right)R + \frac{c_R}{6}\left(\eta-\frac{1}{2}\right)^2 \ , \quad
	R^\eta = R + \frac{c_R}{3}\left(\eta-\frac{1}{2}\right) \ , \quad  Q^\eta = G^-_{-\eta}\;,
\ee
which form the subalgebra
\be
	[2\bar{L}_0^\eta,Q^\eta] = 2\eta Q\;,  \qquad [R^\eta,Q^\eta] = Q^\eta\;,
\ee
and
\be
	\{Q^\eta,(Q^\eta)^\dagger\} = 2\bar{L}_0^\eta - 2\eta R^\eta + \frac{c_R}{3}(\eta^2-\frac{1}{4}) \;,
\ee
where $c_R$ is the right-moving central charge. We refer the reader to~\cite{Schwimmer:1986mf} and references therein for full details of the $\mathcal{N}=2$ superconformal algebra. 

The spectral flow parameter $\eta$ interpolates between the ``R sector" at $\eta=0$, and the ``NS sector" at $\eta=1/2$. Fermions in the R sector have periodic boundary conditions in the $J$-direction in radial quantization, while those in the NS sector are anti-periodic. Clearly, the Hilbert space in radial quantization depends on the parameter $\eta$. We find it informative to keep the parameter $\eta$ and specialize to the NS sector by setting $\eta = 1/2$ at the end of the computation.

The superconformal index is defined as
\be
	\mathcal{I} = {\rm Tr}_{\mathcal{H}_\eta}(-1)^F q^{L_0} a^f\;,
\ee
where $2L_0 = \Delta+J$, and $f$ are Cartan generators of any flavor symmetry, and $q=e^{2\pi \im \tau}$ and $a= e^{2\pi \im u}$ are the corresponding fugacities. The trace is taken over the subspace $\mathcal{H}_\eta$ of the Hilbert space in radial quantization with spectral parameter $\eta$ and annihilated by $Q^\eta$. Using the BPS condition, the index can be rephrased in a rather different form as
\be\label{eq:2d-superconformal-index}
	\mathcal{I} =  {\rm Tr}_{\mathcal{H}_\eta}(-1)^F q^{J+\frac{R}{2}} a^f\ ,
\ee
which turns out to be useful to identify the equivariant parameters for the corresponding symmetries.

In our definition of the superconformal index, we have parametrized the fugacities in the way that is most commonly used in the literature. To conform with the notation used throughout the rest of the paper, we can alternatively write $ 2\pi \im \tau =  -\beta$ and $ 2\pi \im u = -\beta u'$. This will become important when we make contact with the equivariant integral of the anomaly polynomial.

\subsection{Path integral evaluation}

If the SCFT in question admits a UV Lagrangian, the superconformal index admits a path integral formulation on a torus of complex structure $\tau$, which has been evaluated using supersymmetric localization in~\cite{Benini:2013nda,Benini:2013xpa} (see also \cite{Gadde:2013dda}).
The torus is parametrized by a holomorphic coordinate $w=\sigma_1+\tau\sigma_2$ with two periodic real variables $\sigma_1 \sim \sigma_1+2\pi$ and $\sigma_2 \sim \sigma_2+2\pi$. Thus $w$ is periodic with periodicity $w\sim  w + 2\pi \sim w + 2\pi \tau$. We regard the $\sigma_1$ and $\sigma_2$ as ``space" and ``time" coordinates respectively.

The path integral is defined with boundary conditions of the fields along the spatial circle $\sigma_1$. As usual we give all bosonic fields $\Psi_B$ periodic boundary condition. On the other hand, the boundary conditiond for fermionic fields $\Psi_F^\pm$ depend on the chirality $\pm$ and the spectral parameter $\eta$:
\begin{align}
	\Psi_B(\sigma_1+2\pi,\sigma_2) &= \Psi_B(\sigma_1,\sigma_2) \ , \nn \\
	\Psi_F^\pm(\sigma_1+2\pi,\sigma_2) &= e^{\pm2\pi \im \eta} \Psi_F^\pm(\sigma_1,\sigma_2) \ .
\end{align}
In addition the boundary conditions along the time circle $\sigma_2$ are twisted by the flavor chemical potentials.

Let us consider a 2d $\mathcal{N}=(0,2)$ theory with gauge symmetry $G$ and flavor symmetry $F$ together with chiral and Fermi multiplets transforming in representations $\cR^{\mathrm{cm}}$ and $\cR^{\mathrm{fm}}$ respectively. In order to simplify the computation in what follows, we temporarily turn off the chemical potentials for the flavor symmetry $F$. We will also set the $R$-charge of chiral and Fermi multiplets to zero. Both of these parameters can easily be reinstated at the end of the computation.

With these assumptions, the Lagrangians for the chiral and the Fermi multiplets are given by (see for example~\cite{Benini:2013nda})
\begin{align}\label{eq:2d-lagrangian} 
	\mathcal{L}^\cm &= -4 \bar\phi D_{w}D_{\bar{w}}\phi+ \bar\phi(F_{12}+\im D)\phi + 2\bar\psi^-D_{w}\psi^- -\frac{\bar\tau\eta}{\tau_2}\bar\psi^-\psi^- - \bar\psi^-\lambda^+ \phi + \bar\phi \bar\lambda^+ \psi^- \ ,\nn \\
	\mathcal{L}^\fm &= -2\bar\psi^+D_{\bar{w}} \psi^+ + \bar{E}E + \bar{G}G + \bar\psi^+\psi^-_E -\bar\psi^-_E\psi^+ \;,
\end{align}
while the vector multiplet Lagrangian is
\be
	\mathcal{L}^\vm = {\rm Tr}\left[F_{12}^2+D^2-2\bar\lambda^+ D_{\bar{w}}\lambda^+ -\frac{\tau\eta}{\tau_2}\bar\lambda^+\lambda^+\right]\;,
\ee
where
\be
	D_w = \partial_w - \im A_w + \frac{u}{2 \tau_2} f \ .
\ee
The full action is then invariant under
the supersymmetry variation
\begin{align}
	\delta\phi &= - \im\bar\epsilon^+\psi^- \,, \quad \delta\psi^- = 2\im\epsilon^+D_{\bar{w}}\phi \ , \nn \\
	\delta \bar\phi &= -\im\epsilon^+\bar\psi^- \,, \quad \delta\bar\psi^- = 2\im\bar\epsilon^+D_{\bar{w}}\bar\phi \ ,
\end{align}
for the chiral multiplet $(\phi,\psi^-)$ and 
\begin{align}
	\delta\psi^+ &= \bar\epsilon^+ G+ \im\epsilon^+ E \,, \quad \delta G = 2\epsilon^+ D_{\bar{w}}\psi^+ - \epsilon^+ \psi^-_E \ , \nn \\
	\delta\bar\psi^+ &= \epsilon^+ \bar{G} + \im \bar\epsilon^+ \bar{E} \,, \quad \delta \bar{G} = 2\bar\epsilon^+D_{\bar{w}} \bar\psi^+ - \bar\epsilon^+ \bar\psi^-_E \ ,
\end{align}
for the Fermi multiplet $(\psi^+,G)$, and
\begin{eqnarray}
	\delta A_w &=& \tfrac{1}{2}(\epsilon^+\bar\lambda^+ - \bar\epsilon^+\lambda^+) \ , \quad
	\delta \bar\lambda^+ = -\im \bar\epsilon^+(F_{12}-\im D) \ , \quad \delta(F_{12}-\im D) = 2\im D_{\bar{w}}\left(\epsilon^+ \bar\lambda^+ \right) \ , \cr
	\delta A_{\bar{w}} &=& 0 \ , \quad\delta \lambda^+ = \im\epsilon^+(F_{12}+ \im D) \ , \quad  \delta(F_{12}+\im D) = -2iD_{\bar{w}} \left(\bar\epsilon^+ \lambda^+ \right)
\end{eqnarray}
for the vector multiplet $(A_\mu,\lambda^+,D)$.
Here $\psi^-_E = \sum_i \psi^-_i \frac{\partial E(\phi_i)}{\partial \phi_i}$ and the $(\phi_i,\psi^-_i)$'s are chiral multiplets.
We should give the boundary conditions for the supersymmetry parameters and the fermion $\lambda^+$ in the vector multiplet as
\begin{equation}
	\epsilon^\pm(\sigma_1+2\pi,\sigma_2) =e^{\mp2\pi \im\eta}\epsilon^\pm(\sigma_1,\sigma_2) \ , \qquad \lambda^+(\sigma_1+2\pi,\sigma_2) = e^{-2\pi \im\eta}\lambda^+(\sigma_1,\sigma_2)\;,
\end{equation}
so that they are compatible with the supersymmetry variation rules.
Note that the chiral multiplet has a nontrivial fermion mass term proportional to the parameter $\eta$ in the above Lagrangian, but this term can be absorbed by background gauge fields of $U(1)_R$ and flavor symmetries. 

The Lagrangian above is known to be Q-exact and therefore we can use it as a deformation term for localization. 
The 1-loop determinant of this Lagrangian around the saddle points will then yield the exact partition function. See~\cite{Benini:2013nda,Benini:2013xpa} for details. 

To compute the 1-loop determinants we first expand the scalar and fermion fields in terms of their Fourier modes as
\begin{align}
	\phi(w,\bar{w}) &= \sum_{m,n\in \mathbb{Z}} c_{m,n} e^{\im m \sigma_1 - \im n \sigma_2} = \sum_{m,n\in \mathbb{Z}} c_{m,n} e^{-\frac{n+\bar\tau m}{2\tau_2}w + \frac{n+\tau m}{2\tau_2}\bar{w}} \ ,\nn \\
	\psi^+(w,\bar{w}) &= \sum_{m,n\in \mathbb{Z}} b_{m,n}^+ e^{\im\eta\sigma_1}e^{\im m \sigma_1 - \im n \sigma_2} = \sum_{m,n\in \mathbb{Z}} b_{m,n}^+ e^{-\frac{\bar\tau w-\tau\bar{w}}{2\tau_2}\eta} e^{-\frac{n+\bar\tau m}{2\tau_2}w + \frac{n+\tau m}{2\tau_2}\bar{w}} \ , \nn \\
	\psi^-(w,\bar{w}) &= \sum_{m,n\in \mathbb{Z}} b_{m,n}^- e^{-\im\eta\sigma_1}e^{\im m \sigma_1 - \im n \sigma_2} = \sum_{m,n\in \mathbb{Z}} b_{m,n}^- e^{\frac{\bar\tau w-\tau\bar{w}}{2\tau_2}\eta} e^{-\frac{n+\bar\tau m}{2\tau_2}w + \frac{n+\tau m}{2\tau_2}\bar{w}} \ , \nn \\
	\lambda^+(w,\bar{w}) &= \sum_{m,n\in \mathbb{Z}} \tilde{b}_{m,n}^+ e^{-\im\eta\sigma_1}e^{\im m \sigma_1 - \im n \sigma_2} = \sum_{m,n\in \mathbb{Z}} \tilde{b}_{m,n}^+ e^{\frac{\bar\tau w-\tau\bar{w}}{2\tau_2}\eta} e^{-\frac{n+\bar\tau m}{2\tau_2}w + \frac{n+\tau m}{2\tau_2}\bar{w}} \ .
\end{align}
One can easily check that this expansion respects the boundary conditions along $\sigma_1$ and $\sigma_2$.
The twisted boundary condition along the time coordinate $\sigma_2$ can be implemented by turning on the background holonomy for the flavor symmetry.

With this at hand the computation of the 1-loop determinant is straightforward.
For the chiral multiplet, we find
\begin{align}
	Z^\cm &= \prod_{\rho\in \cR} \prod_{m,n\in\mathbb{Z}} \frac{n+\bar\tau m-\langle z,\rho\rangle}{\left(n+\bar\tau m-\langle u,\rho\rangle\right)\left(n+\tau m-\langle z,\rho\rangle\right)} \nn \\
	&= \prod_{\rho\in \cR} \prod_{m,n\in\mathbb{Z}}\left(n+\tau m-\langle z,\rho\rangle\right)^{-1} \, ,
\end{align}
where $z$ denotes the gauge holonomy. For the Fermi and vector multiplets, we find
\begin{equation}
\begin{split}
	Z^\fm &= \prod_{\rho\in \cR}  \prod_{m,n\in\mathbb{Z}}\left(n+\tau m+\tau\eta-\langle z,\rho\rangle\right)\;, \\
	 Z^\vm &= \prod_{e\in \Delta}  \prod_{m,n\in\mathbb{Z}}\left(n+\tau m-\langle z,e\rangle\right)'\ ,
\end{split}
\end{equation}
where the prime on the infinite product in $Z^\vm$ indicates that the zero modes at $m=n=0$ for the Cartan elements are absent.

The results take the form of infinite products, which need to be regularized. We will employ the two-step regularization scheme introduced for the 4d $S^1 \times S^3$ path integral in~\cite{Assel:2015nca}. When applied to the 2d computation this regularization method treats the two Kaluza-Klein towers of modes along $\sigma_1$ and $\sigma_2$ separately.
Thus we expect that this regularization is compatible with the supersymmetric localization, but we will not attempt to prove this here.

We first regularize the infinite product over the KK-modes $m$ along the spatial circle using $\zeta$-function regularization. The result for the chiral multiplet is simply 
\be
\begin{aligned}
	Z^\cm & = \prod_{\rho\in \cR^\cm}\prod_{n\in \mathbb{Z}} \left[  \Gamma_1\Big(\frac{n-\langle z,\rho \rangle}{\tau}\Big|1\Big)\Gamma_1\Big(1-\frac{n-\langle z,\rho \rangle}{\tau}\Big|1\Big) \prod_{m\in \mathbb{Z}}\frac{1}{\tau} \right] \\
	& = \prod_{\rho\in \cR^\cm}\prod_{n\in \mathbb{Z}} \frac{e^{-\pi \im(\frac{1}{2}-\frac{n-\langle z,\rho \rangle}{\tau})}}{1-e^{2\pi \im\frac{n-\langle z,\rho \rangle}{\tau}}}\;,
\end{aligned}
\ee
where the second equality is obtained from the identity in~(\ref{eq:gamma-identity})~\footnote{We also regularize the infinite product $\prod_{m,n\in\mathbb{Z}}1/\tau$ using $\zeta$-function regularization such as 
\begin{equation}
	\prod_{m\in \mathbb{Z}}x=x\Big(\prod_{m>0}x\Big)^2 = x\, e^{2\ln x\cdot \zeta(0)}=x\, e^{-\ln x} = 1 \ ,
\end{equation}
for any nonzero constant $x$.}.

Using the eta and theta functions defined in Appendix~\ref{app:2} and their modular properties, we can rewrite this 1-loop determinant as follows:
\begin{align}
\label{eq:2d-1-loop-chiral}
Z^\cm 
& =\prod_{\rho\in \cR^\cm} e^{ \pi \im(-\frac{1}{2} - \langle z,\rho\rangle^2)/\tau} \frac{\eta(\tau)}{\theta_1\left(\tau\big|\langle z,\rho\rangle\right)}  \cr
&= e^{2\pi \im\tau E^{\rm cm}}\prod_{\rho\in \cR^\cm}\prod_{n\ge1}^\infty\left(1-e^{2\pi \im\langle z,\rho\rangle}q^n\right)^{-1}\left(1-e^{-2\pi \im\langle z,\rho\rangle}q^{n-1}\right)^{-1} \, ,
\end{align}
with
\be
E^{\rm cm} = -\sum_{\rho\in \cR^\cm} f\left[\langle z/\tau ,\rho\rangle+ \frac{1}{2}\right] \, ,
\ee
where we define the function
\be
\qquad f[z] = \frac{z^2}{2} - \frac{1}{24}  \, .
\ee
Similarly, we regularize the Fermi multiplet 1-loop determinant as
\begin{align}\label{eq:2d-1-loop-fermi}
	Z^\fm &= e^{2\pi \im\tau E^{\rm fm}}  \prod_{n\ge1}^\infty \prod_{\rho\in \cR^\fm}(1-e^{2\pi \im \langle z,\rho \rangle}q^{n-\eta})(1-e^{-2\pi \im \langle z,\rho \rangle}q^{n-1+\eta}) \ , \cr
	E^\fm &= \sum_{\rho \in \cR^\fm} f\left[\langle z/\tau,\rho \rangle +\left(\frac{1}{2}-\eta\right)\right] \;,
 \end{align}
and the vector multiplet determinant as
\begin{align}\label{eq:2d-1-loop-fermi}
	Z^\vm &= e^{2\pi \im\tau E^\vm} \prod_{n\ge1}^\infty (1-q^{n})^{2r_{\mathfrak{g}}} \prod_{e \in \Delta^\pm} (1-e^{2\pi \im \langle z,e \rangle}q^{n})(1-e^{-2\pi \im \langle z,e \rangle}q^{n-1}) \ , \cr
	E^\vm &= \sum_{e \in \Delta} f\left[\langle z/\tau,e \rangle +\frac{1}{2}\right] \;,
 \end{align}
where $r_{\mathfrak{g}}$ is the rank of the gauge group. The prefactors $E^{\rm cm},E^{\rm fm}$ and $E^{\rm vm}$ are the contributions to the supersymmetric Casimir energies from the corresponding multiplets. Note  that the spectral parameter $\eta$ does not appear in the results for the vector and chiral multiplets, whereas it remains in the determinant for the Fermi multiplet, as expected.

As a preliminary observation, let us consider the supersymmetric Casimir energies of a free chiral multiplet and a free Fermi multiplet. We find,
\begin{align}\label{eq:Casimir-2d-chiral-fermion}
	 E^{\rm cm} = -\frac{1}{12} \ , \qquad
	 E^{\rm fm}(\eta) = \frac{1}{12} - \frac{\eta(1-\eta)}{2}  \ .
\end{align}
The first equation reproduces the expected vacuum energy for a chiral multiplet. The result for a Fermi multiplet depends on the spectral parameter $\eta$. For Ramond $(\eta=0)$ and Neveu-Schwarz $(\eta=\frac{1}{2})$ sectors, the expected vacuum energies are
\begin{equation}
	{\rm R}\ :\   E^{\rm fm} = \frac{1}{12}  \qquad \ { \rm NS} \ :\ E^{\rm fm} = -\frac{1}{24} \ ,
\end{equation}
which agree with the second formula in~(\ref{eq:Casimir-2d-chiral-fermion}) at $\eta=0$ and $\eta=1/2$.

Let us now return to our gauge theory and reinstate the flavor chemical potentials and non-zero R-charges. At this point we restrict ourselves to the NS sector and so set $\eta = 1/2$. The contributions from chiral, Fermi and vector multiplets, are then
\bea
E^\cm & = -\sum_{(\rho,\rho') \in \cR^\cm} f\left[\langle z/\tau ,\rho\rangle+\langle u /\tau ,\rho'\rangle + \frac{R_{\rho,\rho'}^\cm+ 1}{2}\right] \;,\\
E^\fm &= \sum_{(\rho,\rho') \in \cR^\fm} f\Big [\langle z/\tau,\rho \rangle + \langle u /\tau ,\rho'\rangle + \frac{R_{\rho,\rho'}^\fm}{2} \Big] \;,\\
E^\vm &= \sum_{e \in \Delta} f\left[\langle z/\tau,e \rangle +\frac{1}{2}\right] \;.
\label{eq:Es2d}
\eea
As we discuss in more detail below, in a consistent theory the sum
\be
E \, = \,  E^\cm+E^\fm+E^\vm \;,
\label{eq:Esum}
\ee
is independent of the gauge chemical potential $z$ and gives the total supersymmetric Casimir energy. We now want to compare this to the equivariant integral of the anomaly polynomial.

The anomalies in two dimensions are captured by a four-form anomaly polynomial $A_4$. For a complex left-moving Weyl fermion transforming in a representation $\cR$ of the group $K$, the anomaly four-form is given by
\begin{equation}
	A_4 =  \left[\hat{A}(TM) \cdot {\rm Tr}_\cR\left(e^{F}\right)\right]_4 = \frac{{\rm Tr}_\cR (F^2)}{2} - \frac{p_1(TM)}{24} \ ,
\end{equation}
where $\hat{A}(TM)$ is the A-roof genus of a two-manifold $M$ with a first Pontryagin class $p_1(TM)$, and
$F$ is the field strength for the group $K$. A right-moving Weyl fermion comes with the same anomaly four-form but with overall negative sign, i.e. $A_4^L=-A_4^R = A_4$.

The non-compact scalar $\phi$ in the chiral multiplet minimally coupled to the gauge field as in~(\ref{eq:2d-lagrangian}) has no holomorphic current, so that it does not contribute to the `t Hooft anomaly. Moreover, $\phi$ has equal central charges $c_L=c_R$ and thus it does not contribute to the gravitational anomaly. Therefore we only need to take into account fermion contributions both for chiral and fermi as well as vector multiplets. They are
\begin{itemize}
\item From the chiral multiplets, we have a right-moving fermion with $R$-charge $R_{\rho\rho'}^\cm+1$ for each weight $(\rho,\rho')\in \cR^\cm$.
\item From the Fermi multiplets, we have a left-moving fermion with $R$-charge $R_{\rho,\rho'}^\fm$ for each weight $(\rho,\rho')\in \cR^\cm$.
\item From the vector multiplet, we have a left-moving fermion with $R$-charge $+1$ for each root  of $G$.
\end{itemize}

We will evaluate the equivariant integral of the anomaly four-form on $\mathbb{R}^2$.
One can identify the equivariant parameters for the symmetries with the chemical potentials in the superconformal index~(\ref{eq:2d-superconformal-index}) as in Table~\ref{tb:2d-equivariant-parameters}. Summing the contributions from the fermions listed above, it is straightforward to reproduce the contributions to the supersymmetric Casimir energy in equation~\eqref{eq:Es2d}.

\medskip

\begin{table}[h]
\centering
\begin{tabular}{ | c | c | c|} \hline
  \quad $U(1)_J$ \quad & \quad $U(1)_R$ \quad & \quad $K$ \quad  \\ \hline
  \quad  $\omega=1$ \quad & \quad $\frac{\omega}{2}=\frac{1}{2}$ \quad & \quad $u'=\frac{u}{\tau}$ \quad  \\ \hline
\end{tabular}
\caption{Equivariant parameters from 2d $\cN=(0,2)$ superconformal index.}
\label{tb:2d-equivariant-parameters}
\end{table}

The anomaly polynomial also encodes the quadratic and mixed gauge 't Hooft anomalies. Correspondingly, the putative supersymmetric Casimir energy can include quadratic and linear terms in the holonomy $z$ of the gauge fields. If these terms were present, the periodicity $z \sim z + 1 \sim z + \tau$ will be violated and the path integral would be ill-defined.
To have a consistent theory, the quadratic and mixed gauge 't Hooft anomalies should vanish. This involves the correct assignment of R-charges for the matter multiplets, which can be achieved by {\it c}-extremization \cite{Benini:2012cz,Benini:2013cda}. Then, in a consistent theory, the supersymmetric Casimir energy depends only on the background flavor holonomy and can be pulled outside of the gauge holonomy integral.

\subsection{General formula}
\label{sec:gen2d}

We now want to write a general expression for the supersymmetric Casimir energy of any 2d $\cN=(0,2)$ SCFT. Let us unpack the flavor symmetry into Abelian and simple factors, $F = \prod_b F_b\times \prod_I U(1)_I$. Then the generic form of the four-form anomaly polynomial is
\be
\label{2D anomaly form}
\begin{aligned}
A_{4} = & \frac{k_{RR}}{2} c_{1}(\cF_{R})^{2}
+\sum_I \frac{k_{RI}}{2} c_1(\cF_R)c_1(\cF_I)
+\sum_{I,J}\frac{k_{IJ}}{2} c_{1}(\cF_I)c_{1}(\cF_J) \\
& +\sum_a k_a \, ch_2(\cF_a)
-\frac{k}{24} p_{1}(TM) \, \,.
\end{aligned}
\end{equation}
The anomalies coefficients $k$, $k_{RR}$, $k_{RI}$ and $k_{IJ}$ are defined directly in the SCFT by correlation functions of the appropriate currents, see for example~\cite{Benini:2012cz,Benini:2013cda}. The quadratic gravitational and R-symmetry anomalies are related to the left and right-moving central charges by $k = c_L - c_R$ and $k_{RR} = - 3c_R$ respectively.

Let us denote the fugacities for the Abelian flavor symmetries $U(1)_I$ by $e^{-2\pi \im \tau m_I}$ and those of the simple factors by $e^{-2\pi \im \tau m_a}$ (valued in the Cartan subalgebra of $F$). After equivariant integration of $A_4$, we find that the supersymmetric Casimir energy of a general 2d $\cN=(0,2)$ SCFT is
\be
\label{eq:E2dgen}
E = \frac{1}{8} k_{RR} + \frac{1}{4} \sum_I k_{RI} m_I + \frac{1}{2}\sum_{I,J} k_{IJ} m_Im_J + \frac{1}{2} \sum_a k_a \la m_a , m_a \ra  - \frac{1}{24} k \;.
\ee
In a Lagrangian theory
\bea
k  = \mathrm{Tr}_f(\gamma) \qquad
k_{RI}  = \mathrm{Tr}_f(\gamma R\, q_I) \qquad
k_{IJ} = \mathrm{Tr}_f(\gamma q_I q_J) \qquad
k_a = \mathrm{Tr}_f(\gamma T_aT_a)
\eea
where $R$ is the superconformal R-charge, $q_I$ are the charges with respect to $U(1)_I$, $T_a$ are the Cartan generators of $F_a$, and the traces are over chiral fermions and $\gamma$ is the chirality operator: $\gamma = +1$ for a left-moving fermion and $\gamma = -1$ for a right-moving fermion. These are the standard 't Hooft anomalies from bubble diagrams. In a Lagrangian theory, it is straightforward to show that the result in \eqref{eq:E2dgen} agrees with the expression~\eqref{eq:Esum} we found before for the supersymmetric Casimir energy. 

\section{Discussion}
\label{sec:discussion}

It seems that the most important question is to actually prove, on general grounds, that the supersymmetric Casimir energy in even dimensions is equal to an equivariant integral of the anomaly polynomial. We hope to return to this question in future work. It should be noted that the equivariant integral seems similar to the ``replacement rule" of~\cite{Loganayagam:2012pz,Loganayagam:2012zg,Jensen:2012kj,Jensen:2013rga}. It is tantalizing to explore this connection further. 

Let us mention a few more questions that stem from our work:

\begin{enumerate}
\item We expect that there is a generalization of our results to supersymmetric Casimir energies on manifolds $S^1 \times M$ with $M$ other than $M = S^{D-1}$. Two prominent examples for which this can be explored further are the 4d superconformal index on the Lens spaces $ M =  L(p,q)$, studied in~\cite{Benini:2011nc,Alday:2013rs,Razamat:2013jxa,Razamat:2013opa}, and the partition functions with $M$ some 5d Sasaki-Einstein manifold, analyzed in~\cite{Qiu:2013pta,Qiu:2013aga}.
\item It is usually stated that there are no anomalies in odd dimensions. For three-dimensional theories with at least $\mathcal{N}=2$ supersymmetry however there is a subtle anomaly which was pointed out in \cite{Closset:2012vg,Closset:2012vp}. For these theories on $S^1\times S^2$ there are also prefactors akin to $e^{-\beta E}$, which appear to encode the aforementioned anomalies. It would be interesting to understand whether there exist any characteristic classes whose equivariant integrals reproduce these factors. A preliminary investigation suggests that the Atiyah-Patodi-Singer $\eta$-invariant will play a role. A generalization along these lines to supersymmetric theories in five dimensions will also be interesting.
\item In 2d, the contribution of the supersymmetric Casimir energy to the $T^2$ partition function is crucial to ensure the correct modular properties. It is unclear what is the generalization (if any) of modular invariance to theories in higher dimensions. There are some tantalizing hints from the Cardy formula in four and six dimensions~\cite{DiPietro:2014bca}, which involve the $\beta \to 0$ limit of the partition function (whereas the supersymmetric Casimir energy controls the $\beta \to \infty$ limit).\footnote{See also \cite{Spiridonov:2012ww,Razamat:2012uv} for related work on the modular properties of the 4d superconformal index.} We hope our results may help to elucidate the connection between these limits.
\item Cardy's formula in 2d CFTs relates the leading free energy in the high temperature limit, $\beta \to 0$, to the Virasoro central charge. Analogously, high temperature limits of the superconformal indices in 4d and 6d are conjectured to be fixed by anomalies of SCFTs~\cite{DiPietro:2014bca}. One may wonder if the $\beta\to 0$ asymptotics of the partition function can also be identified with an equivariant integral of characteristic classes. A suggestive observation in this direction is  that the leading term in the 4d superconformal index in the limit $\beta \to0$, as presented in equation (4.5) in~\cite{DiPietro:2014bca}, can be written as the equivariant integral of the 1st Chern classes of the global symmetries. There may also be a similar formula in six dimensions. It is desirable to further understand these results.
\item It is often interesting to study the supersymmetric Casimir energy in the presence of superconformal defects. In the case of 6d $\cN=(2,0)$ SCFTs, the relevant calculations in the ``chiral algebra" limit are presented in~\cite{Bullimore:2014upa} . For general parameters, it may also be possible to extend the 5d partition function computations of Section~\ref{sec:6d} to include defects using results from~\cite{Gaiotto:2014ina,Bullimore:2014awa}. The 4d $\cN=2$ superconformal index in the presence of various kinds of defects has also been computed in~\cite{Gaiotto:2012xa,Gadde:2013dda,Alday:2013kda,Bullimore:2014nla}, which may provide a starting point. Since superconformal defects have an associated anomaly polynomial, there may be a natural extension of our conjecture to this case. 
\item Given the relation between partition functions, indices and anomalies, it should be possible to formulate $a$-maximization in four dimensions~\cite{Intriligator:2003jj} and $c$-extremization in two dimensions \cite{Benini:2012cz,Benini:2013cda} in terms of a statement about supersymmetric partition functions. Since the superconformal R-symmetry in three dimensions is determined by maximizing the partition function of the theory on $S^3$ \cite{Jafferis:2010un}, this will put the ``maximization" principles for SCFTs in two, three and four dimensions on a more equal footing.
\item Since the supersymmetric Casimir energy has an $N^2$ (in 4d) or $N^3$ (in 6d) scaling with the rank of the gauge group it is natural to expect that it should be also accessible by a holographic calculation. This was already discussed to some extent in \cite{Cassani:2014zwa,Assel:2015nca} in four dimensions, but the precise holographic interpretation is not yet clear and deserves further study. It is tantalizing to speculate that there might be a connection between the supersymmetric Casimir energy for $\cN=4$ SYM computed in Section \ref{sec:4d} above and some physical quantity for the Gutowski-Reall black hole \cite{Gutowski:2004ez,Gutowski:2004yv} and its generalizations \cite{Chong:2005hr,Kunduri:2006ek}.
\end{enumerate}

\bigskip

\noindent \textbf{Acknowledgements:} We would like to thank Chris Beem, Francesco Benini, Davide Cassani, Davide Gaiotto, Jaume Gomis, and Phil Szepietowski for useful discussions. We are particularly grateful to Zohar Komargodski for useful discussions, encouragement and comments on the manuscript. The work of NB is supported in part by the starting grant BOF/STG/14/032 from KU Leuven, by the COST Action MP1210 The String Theory Universe, and by the European Science Foundation Holograv Network. NB would like to acknowledge the warm hospitality of Centro de Ciencias de Benasque Pedro Pascual during the final stages of this work. MB gratefully acknowledges support from IAS Princeton through the Martin A. and Helen Choolijan Membership. HK would like to thank the organizers of ``Challenges to Quantum Field Theory in Higher Dimensions'' at Technion, Haifa, and the 2015 Summer Workshop at the Simons Center for Geometry and Physics for their hospitality and support during different stages of this work. The research of HK was supported by the Perimeter Institute for Theoretical Physics. NB and MB are grateful to the Perimeter Institute for the stimulating research environment during the initial stages of this project when they were both postgraduate fellows there. Research at Perimeter Institute is supported by the Government of Canada through Industry Canada and by the Province of Ontario through the Ministry of Economic Development and Innovation.

\appendix

\section{Equivariant characteristic classes and integrals}
\label{app:1}

In this appendix, we will present a brief review on the equivariant characteristic classes and equivariant integration. A more detailed review of this material can be found in \cite{Libine:2007,Tu:2013,Cremonesi:2014dva}. First, consider a compact Lie group $G$ acting on a manifold $M$ and take the maximal torus $T_G$. The equivariant cohomology is then a cohomology defined with the twisted de Rham differential
\be
	d_{\epsilon} = d +\epsilon_a \imath_{X_a} \ ,
\ee
with the equivariant parameters $\epsilon_a$ and the torus elements $X_a\in T_G$. Here $a$ runs over the dimension of the torus action $T_G$. Unlike the ordinary de Rham differential, the twisted differential $d_{\epsilon}$ is no longer nilpotent, but satisfies $d_\epsilon^2 = \epsilon_aL_{X_a}$, where $L_{X_a}$ is the Lie derivative by $X_a$. The $G$ equivariant form $\alpha$ is a cohomology element given by $d_\epsilon\alpha=0$.

As an example, we will analyze the equivariant characteristic classes on a four manifold $\mathbb{R}^4$ with a Lie group $G=U(1)$. Extension to the other symplectic manifold and general Lie groups would be straightforward.
We will introduce equivariant parameters $\omega_{1,2}$ for the $U(1)^2$ rotations on two orthogonal planes in $\mathbb{R}^4$ and $a$ for the $U(1)$ action and define a Lie vector field such as
\be
	X = \omega_1 \left(z_1\partial_{z_1} - z_1^*\partial_{z_1^*}\right)+\omega_2 \left(z_2\partial_{z_2} - z_2^*\partial_{z_2^*}\right) +a L_{U(1)}\ .
\ee
We then define the equivariant de Rham differential with this vector field as follows
\be
	d_\epsilon = d + \imath_X \ .
\ee
The manifold $\mathbb{R}^4$ has a natural symplectic form
\be
	w = dz_1 \wedge dz_1^* + dz_2 \wedge dz_2^* \ ,
\ee
which is $d$-closed, i.e. $dw=0$, but not equivariantly closed by $d_\epsilon$. Using the moment map $\mu= \omega_1 |z_1|^2 + \omega_2 |z_2|^2$, we define the equivariant symplectic form
\be
	e^{-\mu+w} = e^{-\mu}\left(1+w+\frac{w^2}{2!} + \frac{w^3}{3!}+\cdots\right) \ .
\ee
Since $d_\epsilon (\mu+w) = 0 $, this symplectic form is equivariantly closed.

One can construct the equivariant curvature 2-forms using this symplectic form. For example the curvature for the $U(1)$ group can be written as
\be
	F = a\, e^{-\mu+\omega} \ .
\ee
This is a equivariantly closed normalizable 2-form on $\mathbb{R}^4$ and vanishes when $a\rightarrow 0$, as desired. Similarly, the Riemann curvature 2-form associated with the tangent space $TM$ can be written as the following equivariant form
\be
	R_{12} = -R_{21} = \omega_1\, e^{-\mu+w} \ , \quad R_{34}=-R_{43} = \omega_2\, e^{-\mu+w} \ .
\ee
 This is a form-valued $4\times4$ antisymmetric matrix.

We are now ready to perform the integral of differential forms using equivariant localization. The Duistermaat-Heckman (DH) formula tells us that~\footnote{In the main text, we will redefine integrals as $\frac{1}{(2\pi)^d}\int \rightarrow \int$ and omit the $(2\pi)^{-d}$ factors.}
\be
	\frac{1}{(2\pi)^{d}}\int_{M^{2d}} \alpha = \sum_{p} \frac{\alpha|_p}{e(TM)|_p} \ ,
\ee
where $p$ runs over all fixed points of $X$. $\alpha|_p$ is the 0-form component of $\alpha$ evaluated at the $p$'th fixed point and $e(TM)|_p$ is the 0-form component of the equivariant Euler class at $p$.

In the main text we are interested in evaluating equivariant integrals of anomaly polynomials. Let us illustrate how this works for the anomaly 6-form on $\mathbb{R}^4$ 
\be
	\frac{1}{(2\pi)^2}\int A_6 = \frac{1}{(2\pi)^2}\int\left[\hat{A}(R)\cdot Ch(F)\right]_6 = \frac{1}{(2\pi)^2}\int \left[\frac{{\rm Tr}F^3}{6} - \frac{p_1(TM){\rm Tr}F}{24}\right] \ ,
\ee
where $\hat{A}(R)$ is the equivariant A-roof genus associated with the curvature $R$ and $Ch(F)$ is the equivariant Chern character of $F$.
In our case, the vector field $X$ has a single fixed point $p_0$ on $\mathbb{R}^4$ at $z_1=z_2=0$. Hence, by the DH formula, the integral simply reduces to 
\be
	\frac{1}{(2\pi)^2}\int A_6 = \frac{1}{e(TM)|_{p_0}}\left[\frac{{\rm Tr}F^3}{6} - \frac{p_1(TM){\rm Tr}F}{24}\right]_{p_0}  \;, .
\ee
The equivariant Euler class is the Pfaffian of the curvature 2-form $R$, and thus
\be
	e(TM)|_{p_0} = \omega_1\omega_2 \ .
\ee
From the curvature 2-forms $F$ and $R$ defined above, one obtains
\be
	{\rm Tr}F|_{p_0} = a \ , \quad {\rm Tr} F^3|_{p_0} = a^3 \ ,
\ee
and
\be
	p_1(TM)|_{p_0} = -\frac{1}{2}{\rm Tr}R^2|_{p_0} = \omega_1^2+\omega_2^2 \ .
\ee
Plugging these values into the DH formula, we compute the equivariant integral of the anomaly 6-form as
\be
	\frac{1}{(2\pi)^2}\int A_6 = \frac{a^3}{6\omega_1\omega_2} - \frac{(\omega_1^2+\omega_2^2)a}{24\omega_1\omega_2} \ .
\ee

\section{Special functions}
\label{app:2}
In this appendix, we will summarize several special functions used in the paper. The Dedekind eta function is defined as
\begin{equation}
	\eta(\tau) = q^{1/24}\prod_{n=1}^\infty(1-q^n) \ ,
\end{equation}
where $q = e^{2\pi\tau}$. It has the following modular properties:
\begin{equation}
	\eta(\tau+1) = e^{\im\pi/12}\eta(\tau) \ , \quad \eta(-1/\tau) = \sqrt{-\im\tau} \eta(\tau) \ .
\end{equation}
We define the Jacobi theta function as
\begin{equation}
	 \theta_1(\tau|z) = -\im q^{1/8}y^{1/2} \prod_{n=1}^\infty(1-q^n)(1-yq^n)(1-y^{-1}q^{k-1}) \ ,
\end{equation}
with $y=e^{2\pi z}$. The modular properties are
\begin{equation}
	\theta_1(\tau+1|z) = e^{\im\pi/4}\theta_1(\tau|z) \ , \quad \theta_1(-1/\tau|z/\tau) = -\im\sqrt{-\im\tau}e^{\pi \im z^2/\tau} \theta_1(\tau|z) \ .
\end{equation}

The Barnes' multiple zeta function is defined by the series~\cite{2008RuMaS..63..405S}
\be
\zeta_r(s,u;\vec\omega) = \sum_{n_1,\ldots,n_r}^{\infty} \frac{1}{(u+ n_1\omega_1 + \cdots + n_r\omega_r)^s} \;,
\ee
for $s,u\in\mathbb{C}$ and Re$(s)>r$. Choose $\omega_j \in \mathbb{C}$ with $j = 1,\dots, r$ that are linearly dependent over $\mathbb{Z}$. We will often use the notation $\vec\omega = (\omega_1,\ldots,\omega_r)$. We will assume that Re$(\omega_j) \geq 0$ and Im$(\omega_j)>0$. In the context of supersymmetric partition functions these quantities will be complexified squashing parameters.
The function obeys the recursion relation
\be
\zeta_r(s,u+\omega_j;\omega_1,\ldots,\omega_r) - \zeta_r(s,u;\omega_1,\ldots,\omega_r) = - \zeta_{r-1}(s,u;\omega_1,\ldots,\hat \omega_j, \ldots ,\omega_r ) \;,
\ee
with starting point $\zeta_0(s,u;\omega) = u^{-s}$ which allows analytic continuation to Re$(s)\leq r$ except for simple poles at the points $s = 1,\ldots,r$.

The values of the multiple zeta function at $s=0$ are given by the multiple Bernoulli polynomials by the formula
\be
\zeta_r(0,u;\vec\omega) = \frac{(-1)^r}{r!} B_{r,r}(u,\vec\omega)\;,
\ee
where
\be
\frac{x^re^{ux}}{\prod_{j=1}^r (e^{\omega_jx}-1)} = \sum_{n=0}^{\infty} B_{r,n}(u,\omega_1,\ldots,\omega_r) \frac{x^n}{n!} \, .
\ee
The Bernoulli polynomial $B_{r,r}(u,\omega_1,\ldots,\omega_r)$ is a homogeneous polynomial in the variables $u,\omega_1,\ldots,\omega_r$ of degree $r$, divided by the product $\omega_1\ldots\omega_r$. These polynomials play an important role in the relationship between the superconformal index and the partition function on $S^1 \times S^{D-1}$.

Now we define the Barnes' multiple gamma function by 
\be
\Gamma_r(u;\omega) = \exp(\del\zeta(s,u;\omega)/\del s) | _{s=0}\;.
\ee
This obeys the finite difference equation
\be
\Gamma_r(u+\omega_j;\omega_1,\ldots,\omega_r) = \frac{\Gamma(u;\omega_1,\ldots,\omega_r)}{ \Gamma_{r-1}(u;\omega_1,\ldots,\hat \omega_j, \ldots,\omega_r)}\;,
\ee
with initial condition $\Gamma_0(u)= u^{-1}$. 
For example $\Gamma_1(u;\omega) = \omega^{u/\omega} \Gamma(u/\omega) / \sqrt{2\pi \omega}$ is relevant for the hemisphere partition function in two dimensions with radius $\omega = 1/r$.

There are two kinds of infinite product formulae for the Barnes' multiple gamma function that are important for our purposes. Firstly
\be
\frac{1}{\Gamma_r(u;\vec\omega)} = e^{\sum_{j=1}^r\gamma_{r,j}u^j / j!} u \prod_{n_1,\ldots,n_r=0}^{\infty}  \left(1+\frac{u}{ \vec n \cdot \vec \omega} \right) e^{\sum_{j=1}^r(-\frac{u}{ \vec n \cdot \vec \omega} )^j / j}\;,
\ee
where we have used the shorthand notation $\vec n = (n_1,\ldots,n_r)$ and $\gamma_{r,j}$ are some constants like the Euler gamma. The product is understood to omit the zero mode $n_1=\cdots=n_r=0$. This formula arises in evaluating one-loop determinants in the partition function on $S^1 \times S^{D-1}$. There is an important formula involving infinite products
\be\label{eq:gamma-identity}
\prod_{n_1,\ldots,n_r=0}^{\infty}(1-e^{2\pi \im(u+n_1\omega_1+\cdots+n_r\omega_r)}) = \frac{e^{-\im\pi \zeta_{r+1}(0,u;1,\omega) }}{ \Gamma_{r+1}(u;1,\vec\omega) \Gamma_{r+1}(1-u;1,-\vec\omega)}\;,
\ee
which is relevant for relating the partition function on $S^1 \times S^{D-1}$ for  $D$ even to the superconformal index. The appearance of the Bernoulli polynomials in the exponential is of fundamental importance here.

The multiple sine function is also defined as a regularized infinite product~\cite{Ruijsenaars2000107,Friedman2004362}:
\be
	S_r(z|\vec\omega) \sim \frac{\prod_{n_1,\cdots,n_r=0}^\infty (z+\vec\omega\cdot\vec{n}) }
  { \prod_{n_1,\cdots,n_r=1}^\infty (-z+\vec\omega\cdot\vec{n})^{(-1)^r} } \ .
\ee
The multiplet sine functions have another infinite product representation which turns out to be useful in the main text. If $r\ge2$ and Im$(\omega_i/\omega_j)\neq 0$ for $i\neq j$, we can write them as
\bea
	S_r(z|\vec\omega) &= e^{(-1)^r\frac{\pi \im}{r!}B_{rr}(z|\vec\omega)}\prod_{k=1}^r(x_k;\vec{q}_k)_\infty^{(r-2)} \nonumber \\
  &= e^{(-1)^{r-1}\frac{\pi \im}{r!}B_{rr}(z|\vec\omega)}\prod_{k=1}^r(x_k^{-1};\vec{q}_k^{-1})_\infty^{(r-2)} \ ,
\eea
where $x_k=e^{2\pi \im z/\omega_k}, \vec{q}_k = (e^{2\pi \im \omega_1/\omega_k},\cdots,e^{2\pi \im \omega_{k-1}/\omega_k},e^{2\pi \im \omega_{k+1}/\omega_k},\cdots,e^{2\pi \im\omega_r/\omega_k})$.

\bibliographystyle{JHEP}
\bibliography{Indexanomaliesbib}

\end{document}

%% file: preamble.tex
\usepackage{amsfonts}
\usepackage{amscd}
\usepackage{amssymb}
\usepackage{amsmath,bbm}
\usepackage{graphicx}
\usepackage{epsfig}
\usepackage{latexsym}
\usepackage{mathtools}
\usepackage{hyperref}
\usepackage[vcentermath]{youngtab}
    
\def \be  {\begin{equation}}
\def \ee  {\end{equation}}
\def \bea {\begin{equation}\begin{aligned}}
\def \eea {\end{aligned}\end{equation}}
\def \ba  {\begin{eqnarray}}
\def \ea  {\end{eqnarray}}
\def \bb  {\begin{thebibliography}}
\def \eb  {\end{thebibliography}}
\def \lab #1 {\label{#1}}

\newcommand\cF{\mathcal{F}}

\newcommand\cH{\mathcal{H}}
\newcommand\cI{\mathcal{I}}

\newcommand\cN{\mathcal{N}}
\newcommand\cO{\mathcal{O}}

\newcommand\cR{\mathcal{R}}
\newcommand\cS{\mathcal{S}}
\newcommand\cT{\mathcal{T}}

\newcommand\cW{\mathcal{W}}

\newcommand\im{\mathrm{i}}

\newcommand\cm{\mathrm{cm}}
\newcommand\fm{\mathrm{fm}}
\newcommand\vm{\mathrm{vm}}

\newcommand\la{\langle}
\newcommand\ra{\rangle}
\newcommand\del{\partial}

\newcommand\tr{\mathrm{Tr}}

\newcommand{\PE}[1]{ \mathrm{PE}\left[ #1 \right] }

\def\im{{\rm i}}